\documentclass[useAMS,usenatbib,usegraphicx,paper]{mn2e}
\usepackage{longtable}

\def \hcn{HCN}
\def \hco{HCO$^+$}

\title[Dense gas in Seyfert galaxies]
  {Physical properties of dense molecular gas in centres of Seyfert galaxies}
  
  \author[E. Sani et al.]
  {E.~Sani,$^{1,2,}$\thanks{sani@arcetri.astro.it},
  R.~I.~Davies,$^2$ A.~Sternberg,$^3$ J.~Graci{\'a}-Carpio$^{2}$, E. K. S.~Hicks$^{4}$,
  \newauthor  M.~Krips$^{5}$, L. J.~Tacconi$^{2}$, R.~Genzel$^{2}$, B.~Vollmer$^{6}$, E.~Schinnerer$^{7}$, 
  \newauthor S.~Garc{\'i}a-Burillo$^{8}$, A.~Usero$^{8}$, G.~Orban de Xivry$^{2}$    
             \\
  $^1$INAF - Osservatorio Astrofisico di Arcetri, Largo E. Fermi 5, I-50125 Firenze, Italy\\
  $^2$Max-Planck-Institut f\"{u}r extraterrestrische Physik, Postfach 1312, 85741 Garching, Germany\\ 
  $^3$School of Physics and Astronomy, Tel Aviv University, Tel Aviv 69978, Israel\\
  $^4$Department of Astronomy, University of Washington, Seattle, WA 98195-1580, USA\\
  $^5$Institut de Radioastronomie Milim\'{e}trique, 38406 Saint Martin d’Heres, France\\
  $^6$CDS, Observatoire Astronomique de Strasbourg, 11 rue de l'universit{\'e}, 67000 Strasbourg, France\\
  $^7$Max-Planck-Institut f\"{u}r Astronomie, K\"{o}nigstuhl 17, Heidelberg 69117, Germany\\
  $^8$Observatorio Astron{\'o}mico National, Alfonso XII 3, 28014 Madrid, Spain}

\date{Received ...; accepted ...}

\pagerange{\pageref{firstpage}--\pageref{lastpage}} \pubyear{2012}

\def\LaTeX{L\kern-.36em\raise.3ex\hbox{a}\kern-.15em
    T\kern-.1667em\lower.7ex\hbox{E}\kern-.125emX}

\begin{document}

\label{firstpage}

\maketitle

\begin{abstract}
We present new $\sim1\arcsec$ resolution data of the dense molecular gas in the central $50-100$~pc 
of four nearby Seyfert galaxies. PdBI observations of \hcn\ and, in 2 of the 4 sources, simultaneously \hco\ allow us to carefully constrain the dynamical state of the dense gas surrounding the AGN. 
Analysis of the kinematics shows large line widths of 100--200~km/s FWHM 
that can only partially arise from beam smearing of the velocity gradient. 
The observed morphological and kinematic parameters 
(dimensions, major axis position angle, red and blue channel separation, and integrated line width) 
are well reproduced by a thick disk, where the emitting dense gas has a 
large intrinsic dispersion (20--40~km/s), implying that it exists 
at significant scale heights (25--30\% of the disk radius).
To put the observed kinematics in the context of the starburst and AGN 
evolution, we estimate the Toomre Q parameter. We find this is always greater than 
the critical value, i.e. Q is above the limit such that the gas is stable against rapid star formation. 
This is supported by the lack of direct evidence, in these 4 Seyfert galaxies, for on-going star formation close around the AGN.
Instead, any current star formation tends to be located in a circumnuclear ring. 
We conclude that the physical conditions 
are indeed not suited to star formation within the central $\sim100$~pc.
\end{abstract}


\section{Introduction}

The key component of the unification model for active galactic nuclei (AGN) is a 
geometrically thick torus (e.g. \citealt{ant93}, \citealt{urr95})
where gas and dust obscure the primary optical-ultraviolet AGN emission, depending on the 
line-of-sight viewing angle.
The signature of such a torus is a pronounced infrared (IR) peak in the spectral energy 
distribution (e.g. \citealt{elv94}), which is interpreted 
as thermal emission from hot dust heated by the primary radiation (\citealt{bar87}).
In this scenario the obscuring material surrounding the AGN accretion disk has 
a hydrostatic toroidal geometry and is thought to begin at the dust sublimation radius 
and extend for tens, or possibly even hundreds, of parsecs 
\citep{pie92,gra94,nen02,nen08,sch05,sch08,hoe06}.
A smooth continuous distribution of the medium in the torus leads to stability problems. 
Thus it has been suggested that the thick torus should either be clumpy 
\citep{nen02,eli04,fri06},
or be supported by an additional force combined with the thermal pressure.
Nuclear star formation (SF) can play an important role in sustaining its vertical extent, 
e.g. through stellar radiation pressure \citep{tho05,bal08},
supernovae explosions \citep{wad02}, 
or stellar winds \citep{nay07}.
Thus the connection between AGN activity and nuclear SF is an important issue 
related to the torus structure. Indeed nuclear SF happens in all types 
of AGN \citep{cid04}, and there is evidence that enhanced SF 
(reaching starburst intensities) is related to the black hole accretion rate \citep{san10}.

Thanks to the capabilities of the adaptive optics assisted integral field spectroscopy, 
it has recently become possible to study the inner 100~pc of galaxies with a spatial resolution 
of $\sim$10~pc. \cite{dav06} used SINFONI to analyse the molecular gas and stars in NGC~3227, finding the central gas distribution was geometrically and optically thick, analogous to the standard molecular torus. 
Furthermore they found a recent, but no 
longer active, starburst within the tens of parsecs around the AGN that occurred $\sim40$~Myr ago. 
A similar result is found in a larger sample of AGN galaxies, where there appears 
to be a delay of 50--100~Myr between the onset of star formation and the onset 
of AGN activity \citep{dav07}.
On similar scales of tens of parsecs, 
\citet[hereafter H09]{hic09} traced the distribution and kinematics of the warm molecular gas, 
which they associated with the largest scale structures of the torus.
Analysis of the H$_2$~(1-0)\,S(1) emission in these AGN showed that the gas 
has a huge velocity dispersion (50--100 km/s), implying it has a geometrically thick structure 
with a height of the order of the disk radius ($H\sim R$). 
Moreover, based on spatial distribution, column density and kinematics, H09 concluded 
that the molecular gas is spatially mixed with the nuclear stellar population. 

In an attempt to further understand the circumnuclear structure around AGN, we present here \hcn(1-0) and \hco(1-0) 
distributions and kinematics in four nearby Seyfert galaxies observed with the 
\emph{Plateau de Bure Interferometer}~(PdBI)\footnote{The PdBI is managed by the 
IRAM institute, which is supported by INSU/CNRS (France), MPG (Germany) and IGN (Spain).}.
With this array, it is possible to reach a spatial resolution of $\leq 1\arcsec$ at 89~GHz. 
Although this is still one order of magnitude larger than the resolution obtainable at near-IR wavelengths, the millimetre regime offers a number of clear advantages, most importantly the excellent spectral resolution and that the emission is free of extinction.

Our main purpose of this paper is to understand whether the high dispersion seen previously in the warm H$_2$ is related 
to the tracer properties, or whether it is really representative of the bulk gas properties. 
This is an important distinction because the 2.12~$\mu$m H$_2$ line traces just the warm ($\sim 1000$~K) component of the gas, which accounts only for a tiny fraction of the total gas mass ($10^{-6}-10^{-5}$). 
In contrast, the \hcn(1-0) and \hco(1-0) transitions probe the cold ($\sim10-100$K) and dense ($n > 3\times10^4$cm$^{-3}$) 
molecular gas \citep{gra06,pap07} and are thus suitable for our purposes. 
This is the case mainly for two reasons: (\emph{i}) \hcn\ and \hco\ lines trace 
$\sim 100-500$ times denser gas than corresponding (rotational) CO transitions and 
(\emph{ii}) \hcn\ (and \hco) are relatively strong close to the AGN, and weaker elsewhere. 
In fact the \hcn\ (and \hco) to CO intensity ratio can be significantly higher 
in AGN than in starburst or quiescent regions \citep{kri07,gra08,koh08,dav12}.
We note that the X-ray irradiation of the gas by the AGN may affect the nuclear chemistry (i.e. molecular abundances).
For example, models in \cite{bog05} and \cite{mei05} show that the equilibrium abundances depend 
on the ratio of the local gas density and the incident X-ray ionization rate. 
These effects are beyond the scope of this work, 
but are investigated with specific reference to NGC\,3227 by \cite{dav12}.
Despite this, a modified chemistry should have little impact on the molecular gas kinematics. 
The use of two tracers, when available, reduces the risk that the observed kinematics are skewed by 
chemistry affecting the inferred gas mass distribution. 
As such, \hcn\ and \hco\ lines can be used to reliably measure the characteristic kinematics of the 
dense gas in the central region. 
 
In this paper we present \hcn(1-0) observations for NGC~2273, NGC~3227, NGC~4051 and NGC~6951 
together with \hco(1-0) data for NGC~2273 and NGC~4051. 
Obervations and data reduction are discussed in Section~2, and the general 
properties of the molecular gas are presented in Section~3. 
In Section~4 we analyse the dense gas kinematics, and Section~5 is dedicated to the 
evolutionary interpretation. Our conclusions are given in Section~6.

\section{Observations}

\begin{table*}
 \begin{center}
 \begin{tabular}{lcccccc}
 \hline
  Source   & Program & Beam Size & Beam PA & Channel res. & rms &pc scale\\
           &          & $\arcsec\times\arcsec$ &  deg    &   km/s & mJy/beam    & pc/$\arcsec$\\
  \hline
  NGC~2273 & T095         &  1.19 x 0.90       & 38 & 21 & 0.76 &130 \\
  NGC~3227 & S098         &  1.19 x 0.68       & 47 & 50 & 0.65 &85 \\
  NGC~4051 & T095         &  1.07 x 0.72       & 70 & 21 & 0.63 &50 \\
  NGC~6951 & PB67         &  1.42 x 1.09       & 73 & 4  & 1.6   &102 \\
  \hline
  \end{tabular}
   \label{tb:obs}
  \caption{(1) Source name. (2) Observing program. 
  (3) Beam size in A configuration, or A+B configuration for NGC~6951. 
  (4) Position angle of the beam, measured East of North.
  (5) Spectral resolution (after binning). 
  (6) Root mean square uncertainty.
  (7) Spatial scale at the distance of the source.}
 \end{center}
\end{table*}

\begin{table*}
 \begin{tabular}{lccc}
 \hline
  Source   & Flux Density & Major x Minor axis & Pos. angle \\
           &     mJy      & $\arcsec\times\arcsec$ &   deg \\
  \hline
  NGC~2273 & $2.34\pm0.09$& (1.86$\pm$0.05) x (1.13$\pm$0.03)        & 87$\pm3$ \\
  NGC~3227 & $1.5\pm0.3$  &  (1.15$\pm0.06$) x (1.05$\pm0.08$)       & 47$\pm7$ \\
  NGC~4051 & $0.9\pm0.2$  & (1.13$\pm0.07$) x (0.90$\pm0.09$)        & 82$\pm10$\\
  \hline
  \end{tabular}
  \caption{Parameters of the continuum emission for the new observations. (1) Source name. (2) Flux density. 
  (3) Observed major and minor axis diameters. (4) Position angle of the major axis, measured East of North.}
 \label{tb:cont}
\end{table*}

We obtained data using the PdBI (which has 6 antennaes of 15 m diameter) in the extended A configuration (760 m maximum baseline) for two sets of observations, the first relating to NGC~3227 and the second for 
NGC~2273 and NGC~4051. NGC~6951 was observed previously in A+B configuration, and 
the data published by K07.
However, the K07 work deals mostly with the spatial trend of the \hcn/\hco\ ratio.
Our analysis here concentrates instead on the central source kinematics. 
All the data presented here were processed and calibrated using 
the CLIC program in the IRAM GILDAS package\footnote{The GILDAS software is available at\\ http://www.iram.fr/IRAMFR/GILDAS}.
We refer the reader to K07 for all the details concerning the observing conditions for NGC~6951.
The observations of the other targets are described below.

Observations of NGC~3227 were carried out during January and February 2009 for program S098. 
In the 1~mm band, the H$^{13}$CN(3-2) line at 259.01 GHz was observed in a single 
7.6 hr track using the dual polarization receivers. Atmospheric conditions were moderate 
on this night and the system temperature (T$_{sys}$) was about 160 K. In the 3~mm band the 
H$^{12}$CN(1-0) line at 88.63 GHz and the H$^{13}$CN(1-0) line at 86.34 GHz were 
observed simultaneously. This was achieved by observing a 1~GHz bandwidth segment 
across each line using a single polarization for each of the two segments. 
These data were acquired over tracks totalling 14.5~hrs during 3 nights with mediocre and variable atmospheric 
conditions, with up to 4-6~mm water vapour and sometimes strong wind. 
The T$_{sys}$ was 80-100~K. 
The bandpass of the receivers was calibrated by observing the quasar 3C~273 (first and second night) and 
0851+202 (third night). 
Amplitude and phase calibrations were achieved by monitoring 0923+392, 
whose flux densities were determined relative to 3C~273, MWC349 or 0851+202 resepctively.
We applied a spectral binning of 50~km/s to increase the signal to noise. 
The resulting angular beam sizes were $1.19\arcsec \times 0.90\arcsec$ at position angle PA$=38^\circ$ (East of North) 
for the H$^{12}$CN(1-0) line; $1.21\arcsec \times 0.93\arcsec$ at $37^\circ$ for the H$^{13}$CN(1-0) line; 
$0.56\arcsec \times 0.38\arcsec$ at PA$=23^\circ$ for the H$^{13}$CN(3-2).

A detection of the H$^{12}$CN(1-0) line in NGC~3227 was previously reported by \cite{sch00a} at a resolution of $2.4\arcsec$. 
Our more sensitive and higher resolution 
observations resolve the source and detect the 3~mm continuum. 
Neither the 
H$^{13}$CN(1-0) nor the H$^{13}$CN(3-2) lines were detected.

We obtained 3~mm data for the Seyfert nuclei NGC~2273 and NGC~4051 with program T095, executed during December 2009. 
The observations were performed as for NGC~3227 but with the central frequency shifted 
in order to simultaneously observe both \hcn(1-0) at 88.63~GHz and \hco(1-0) at 89.19~GHz. 
NGC~2273 was observed with 2 tracks of 1.5 + 5.5 hr on source. For both tracks, the atmosphere was 
stable, resulting in T$_{sys}= 65~K$ and T$_{sys}= 80~K$ respectively. 
A single track of 7.7 hr was obtained for NGC~4051 with good sky conditions 
(wind $>15$~m/s just at the end of the track) and T$_{sys}= 50~K$.
For NGC~2273, the bandpass was calibrated by observations of 0923+392 (short track) and 
3C454.3 (long track), while 0646+600 and 0716+614 were used as amplitude and phase 
calibrators respectively for the two tracks. 
During the long track, fluxes were calibrated using 3C454.3 and checked on 0716+714 and 0646+600; 
while for the short track fluxes where calibrated by monitoring 0646+600, 0716+714 and 1055+018 with respect to 0923+392. 
For NGC~4051, the bandpass calibrator was 3C273. Phase and amplitude calibrations were achieved by monitoring 
J1224+435 and 1128+385, whose flux densities were determined relative to 0923+392.
The data quality assessment allowed us to apply a 20~km/s 
spectral binning. The final angular beams were: $1.19\arcsec \times 0.68\arcsec$ at PA$=47^\circ$
and $1.07\arcsec \times 0.72\arcsec$ at PA$=70^\circ$ in NGC~2273 and NGC~4051 respectively.
The rms phase noises were between $30^\circ$ and $43^\circ$ at 3~mm, which introduces absolute position errors 
$\leq 0.1\arcsec$ in our observations. But we note that atmospherically or instrumentally induced relative position errors within each dataset are 
significantly smaller than this, so that separation measurements depend instead almost entirely on the centroiding accuracy.

To achieve a high spatial resolution, here we analyse 
uniformly weighted maps. This allows us to avoid contamination by circumnuclear structures, such as the ring emission in NGC~3227 \citep{dav06} and NGC~6951 (K07).
Table~1 lists the observing parameters.  
\section{Results}

\begin{table*}{Molecular Lines}
 \begin{center}
 \begin{tabular}{llcccccc}
 \hline
 Source & Line & Flux & Major x Minor axis & PA         & r/b Sep       & PA$_{r/b}$  & FWHM\\
        &       & Jy~km/s & $\arcsec\times\arcsec$ & deg    & $\arcsec$     & deg         & km/s \\
 \hline
  NGC~2273 & \hcn\ & $1.76\pm0.07$& (1.87$\pm$0.05) x (1.17$\pm$0.03) & 18$\pm$6 & 1.10$\pm$0.02  & 23.2$\pm$0.4       & $181\pm21$\\     
           & \hco\ & $2.13\pm0.08$& (1.94$\pm$0.04) x (1.34$\pm$0.04) & 44$\pm$6           & 0.87$\pm$0.03  & 26.2$\pm$0.8       & $175\pm16$\\
 \hline
  NGC~3227 & \hcn\ & $1.86\pm0.27$& (1.28$\pm$0.06) x (1.04$\pm$0.06) & 47$\pm$2      & 0.38$\pm$0.01 & -37.1$\pm$0.4 & $207\pm34$\\
 \hline
  NGC~4051 & \hcn\ & $0.91\pm0.05$& (1.92$\pm$0.09) x (1.02$\pm$0.05) & 80$\pm$5      & 0.55$\pm$0.02   & 39.4$\pm$0.6   & $74\pm10$\\
           & \hco\ & $2.07\pm0.05$& (1.65$\pm$0.05) x (1.27$\pm$0.04) & 72$\pm$2      & 0.60$\pm0.04$        & 19.6$\pm$0.7     & $90\pm10$ \\
 \hline
  NGC~6951 & \hcn\ & $1.02\pm0.02$& (1.44$\pm$0.03) x (1.10$\pm$0.03) & -80$\pm$3     & 0.60$\pm$0.01  & -50.2$\pm$0.5      & $173\pm29$\\
 \hline
 \end{tabular}
 \caption{Parameters of the molecular emission: (1) source name. (2) line transition: \hcn(1-0) or \hco(1-0). 
 (3) Line flux within the central $3\arcsec$. (4) Observed major and minor axis diameters of the molecular emission. 
 (5) Position angle of the major axis, measured East of North. 
 (6) Separation of the red and blue velocity channel maps (see text for details of the channels summed to produce these). 
 (7) Position angle (east of north) of the vector joining the centroids of the red and blue channel maps. 
 (8) Width of the molecular line (FWHM) from the integrated spectrum.}
 \end{center}
 \label{tb:lines}
\end{table*}

\begin{figure*}
\centerline{
\hbox{
\includegraphics[width=0.3\linewidth,angle=0]{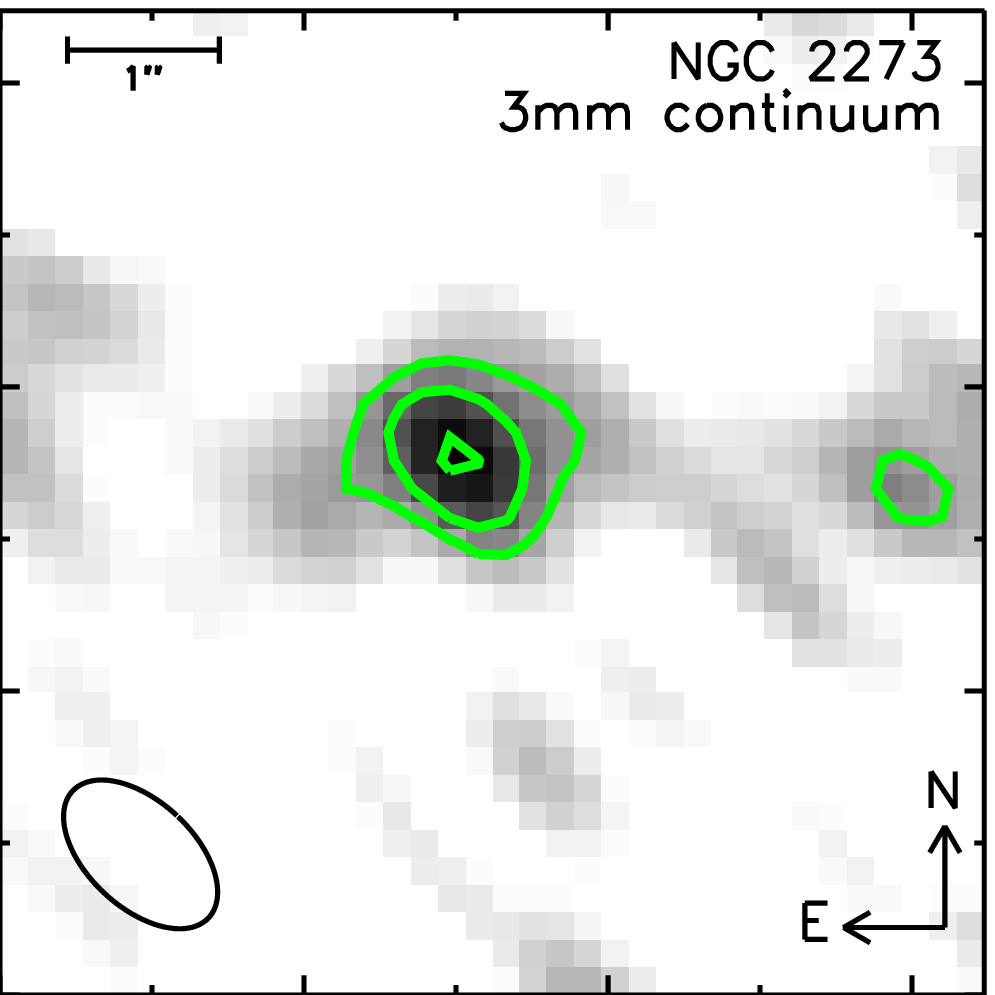}
\includegraphics[height=0.3\linewidth]{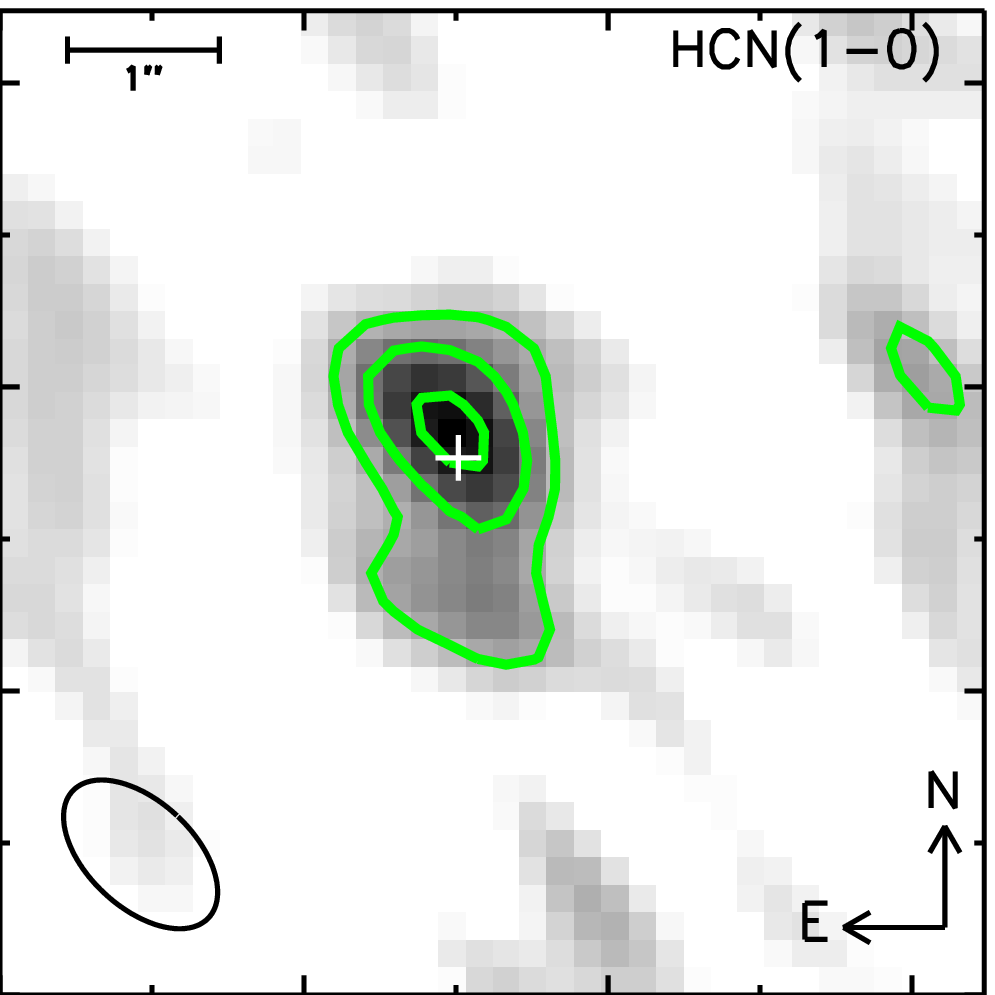}
\includegraphics[height=0.3\linewidth]{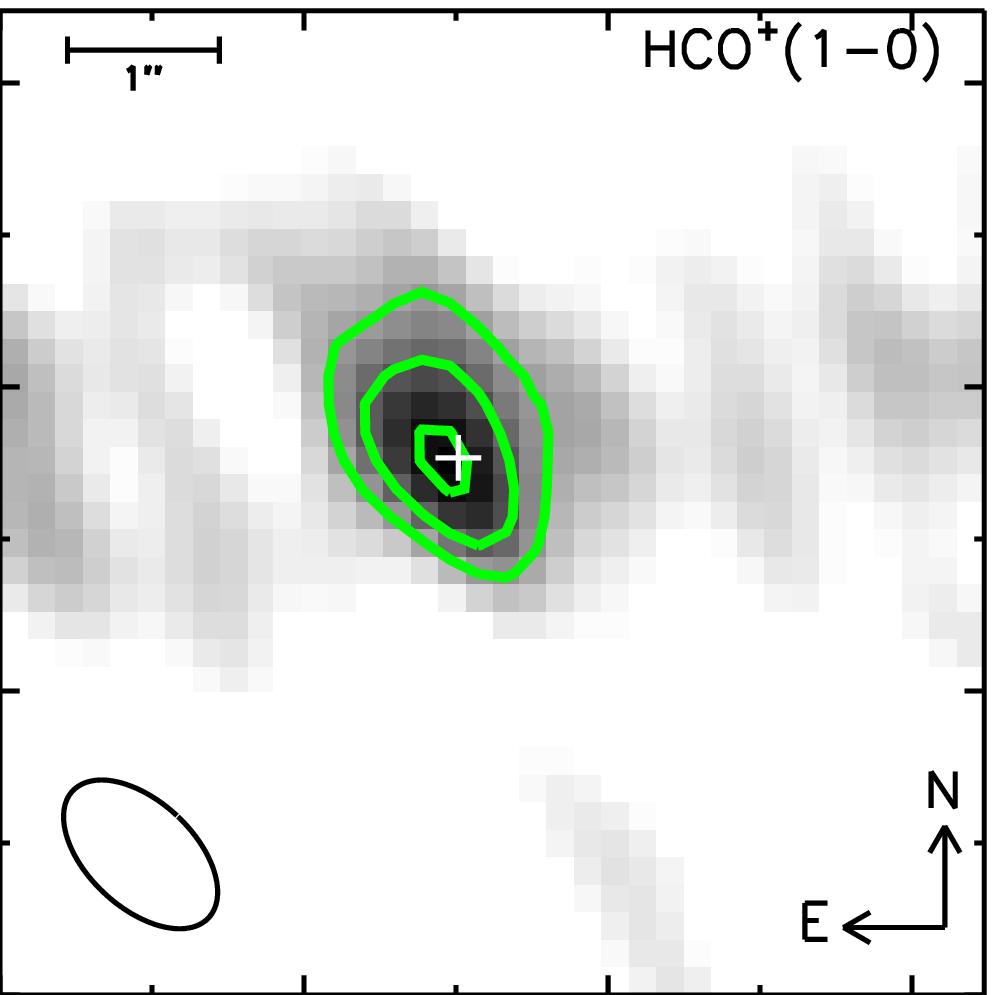}
}
}
\caption{The 
3~mm continuum (left, rms~$=0.04$~Jy~beam$^{-1}$km~s$^{-1}$), 
\hcn\ (middle, rms~$=0.08$~Jy~beam$^{-1}$km~s$^{-1}$) and 
\hco\ (right, rms~$=0.11$~Jy~beam$^{-1}$km~s$^{-1}$) emission maps of NGC~2273. 
The contour levels are at 2, 4, and 6, times their respective noise levels.
The beam size (FWHM) is shown in the lower left of each panel, 
and the white `plus' signs denote the peak of the continuum.
The continuum appears partially resolved, while the line emission is more obviously extended in both \hcn\ and \hco\ lines.}
\label{fg:2273}
\end{figure*}

We first describe the overall properties of our sources, followed by 
a more detailed analysis source by source. 
Figures~\ref{fg:2273}, \ref{fg:3227}, and \ref{fg:4051}, show the continuum,
\hcn, and (when available) \hco\ maps for NGC~2273, NGC~4051 and NGC~3227 respectively. 
For these sources, because the continuum is only weakly detected in single spectral channels, 
we obtain continuum maps by collapsing all the channels with no signal from the molecular transitions 
(i.e. neither \hcn\ nor \hco). The line maps are created by summing those 
channels where the emission is detected and subtracting the continuum. 
In Fig.~\ref{fg:6951} we show the \hcn\ emission for NGC~6951 from K07. 

The 3~mm continuum is detected in 3 of the 4 objects as a single compact (mostly) unresolved source. 
The sizes and flux densities of the continuum sources are given in Table~\ref{tb:cont}.
Similarly, the line emission appears with a centrally concentrated morphology located at the same 
position, but (marginally) resolved. 
We measure the extent of each source both as the projected shape in the \textit{uv} plane and in spatial coordinates. 
Because the data are characterized by a relatively low signal-to-noise ratio (S/N), 
the fitting in \textit{uv} coordinates is performed using a symmetric Gaussian.
This bi-dimensional fitting is performed using the MAPPING program in the GILDAS package, and gives a direct measure 
(i.e. without beam convolution) of the intrinsic source size.
If the source distribution is highly non-circular this may not be 
realistic and so 
this size estimate is used only as a consistency check.
Our primary size measurement is based on a fit to the data in spatial coordinates (i.e. including the cleaned beam). 
Thus, the FWHMs in the two spatial directions represent the \textit{observed} major and minor 
axes of the sources.
We note, however, that the fluxes and their relative uncertainties are integrated within a $3 \arcsec$ aperture around the image center. 
To properly compare the \textit{observed} source dimensions (in spatial coordinates) with \textit{intrinsic}
measurements (in the \textit{uv} plane), it is necessary to  subtract in quadrature the beam shape from the former.
We derive the uncertainties for the observed sizes using Monte Carlo techniques 
by creating $10^4$ realizations of the image, each time adjusting the data 
values with random numbers according to the measured noise and re-fitting the 
Gaussian and background. 
For the two objects in which both lines are observed, the similarity of the spatial parameters 
of the \hcn\ and \hco\ emitting regions given in Table~\ref{tb:lines}, 
verifies that the two molecules are likely to be tracing the same gas components. 

The spectra of the sources are shown in Figures~\ref{fg:sp2273}, \ref{fg:sp3227}, \ref{fg:sp4051} and~\ref{fg:sp6951}. 
These are integrated over all spatial elements within a $3\arcsec$ aperture
 in which the signal exceeds a 0.25~mJy~beam$^{-1}$ threshold. 
 This threshold is significantly lower than the rms given in Table~1 
 and is simply used to increase the S/N without affecting the resulting integrated line widths (which are consistent within the errors, both with and without the threshold).
We derive the uncertainties for the line FWHMs using Monte Carlo realizations of each spectrum, as described above for observed source sizes. 
The most remarkable characteristic is the large line width: $207\pm34$ and $173\pm29$~km/s 
for the \hcn\ line in NGC~3227 and NGC~6951, respectively; and mean FWHMs of $177\pm14$ and $82\pm10$~km/s for NGC~2273 and NGC~4051 respectively. 
Such widths are comparable with those observed for the warm gas traced by the H$_2$~(1-0)\,S(1) line, reported by H09. 
As discussed in the next section, the implied high velocity dispersion is a crucial 
parameter to probe the dense gas distribution.

The spatial sizes, velocity widths, and fluxes of the lines are listed in Table~\ref{tb:lines}. 
We emphasize that the measurements in this section are \textit{observed} quantities. 
To perform an appropriate dynamical analysis of the gaseous structure we need \textit{intrinsic} 
properties, i.e. we have to take into account 
the impact of spatial and spectral beam smearing (see Section~4 for details).

\subsection{NGC~2273}

\begin{figure}
\vbox{
\includegraphics[height=0.6\linewidth]{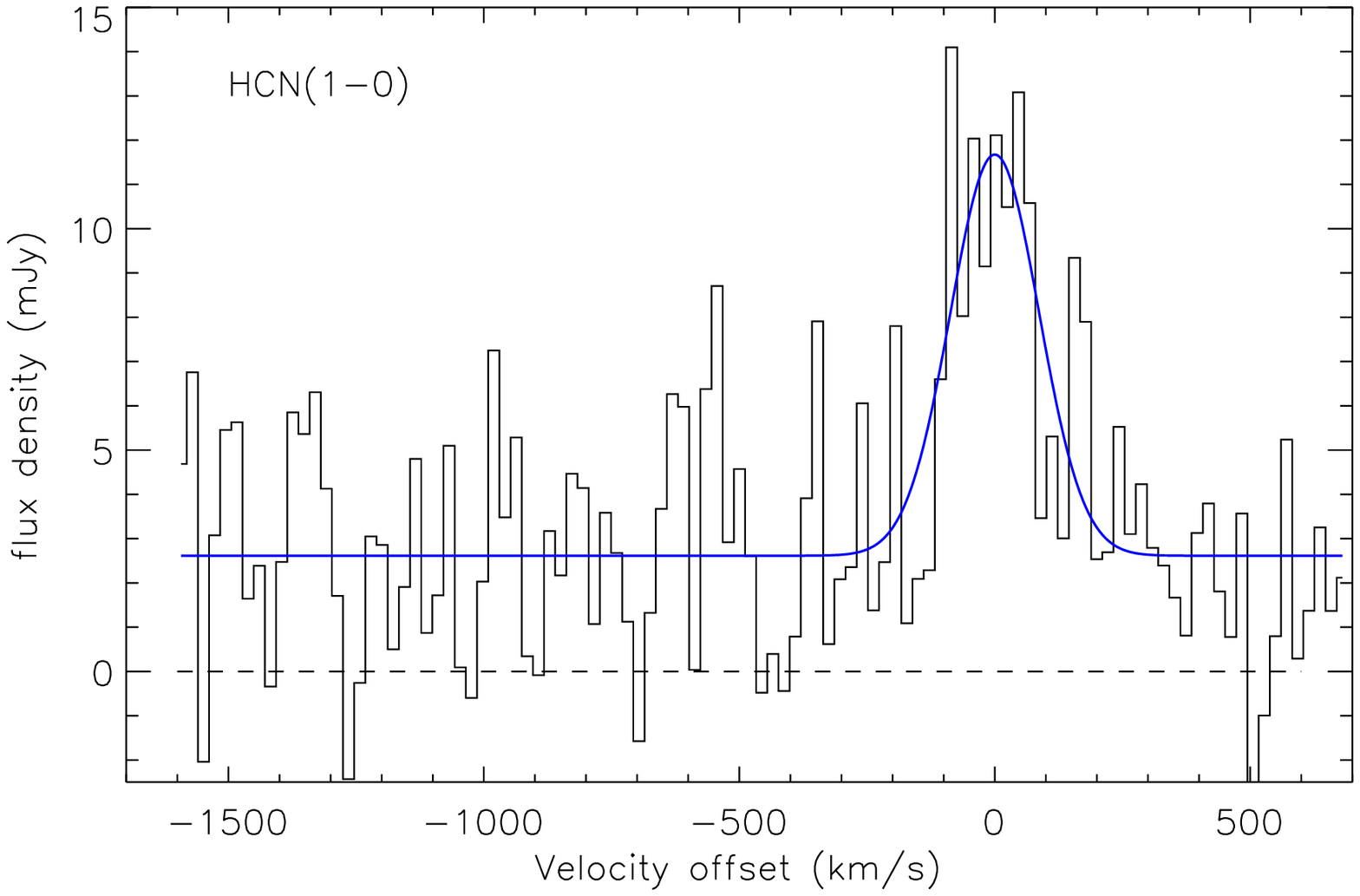}
\includegraphics[height=0.6\linewidth]{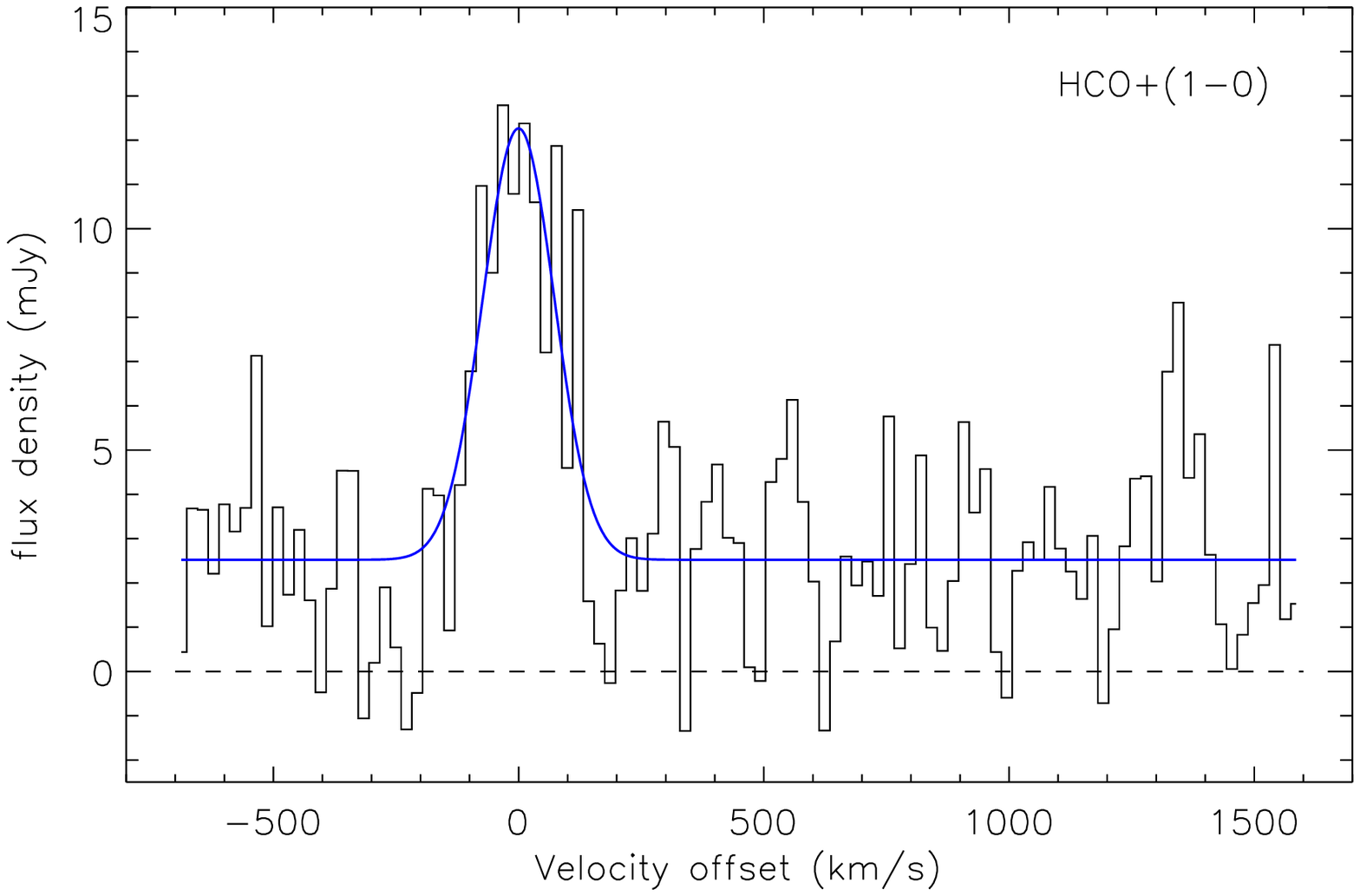}
}
\caption{Integrated spectra of NGC~2273 for the \hcn\ (top), and \hco\ (bottom) lines. 
The blue curve is a fit (Gaussian plus constant term) to the 3~mm continuum and the 
line emission, which can be seen to have a large velocity width (mean FWHM of the two lines is $177\pm14$~km/s). 
The dashed line represents the zero flux level.}
\label{fg:sp2273}
\end{figure}

\begin{figure*}
\centerline{
\hbox{
\includegraphics[height=0.3\linewidth,angle=0]{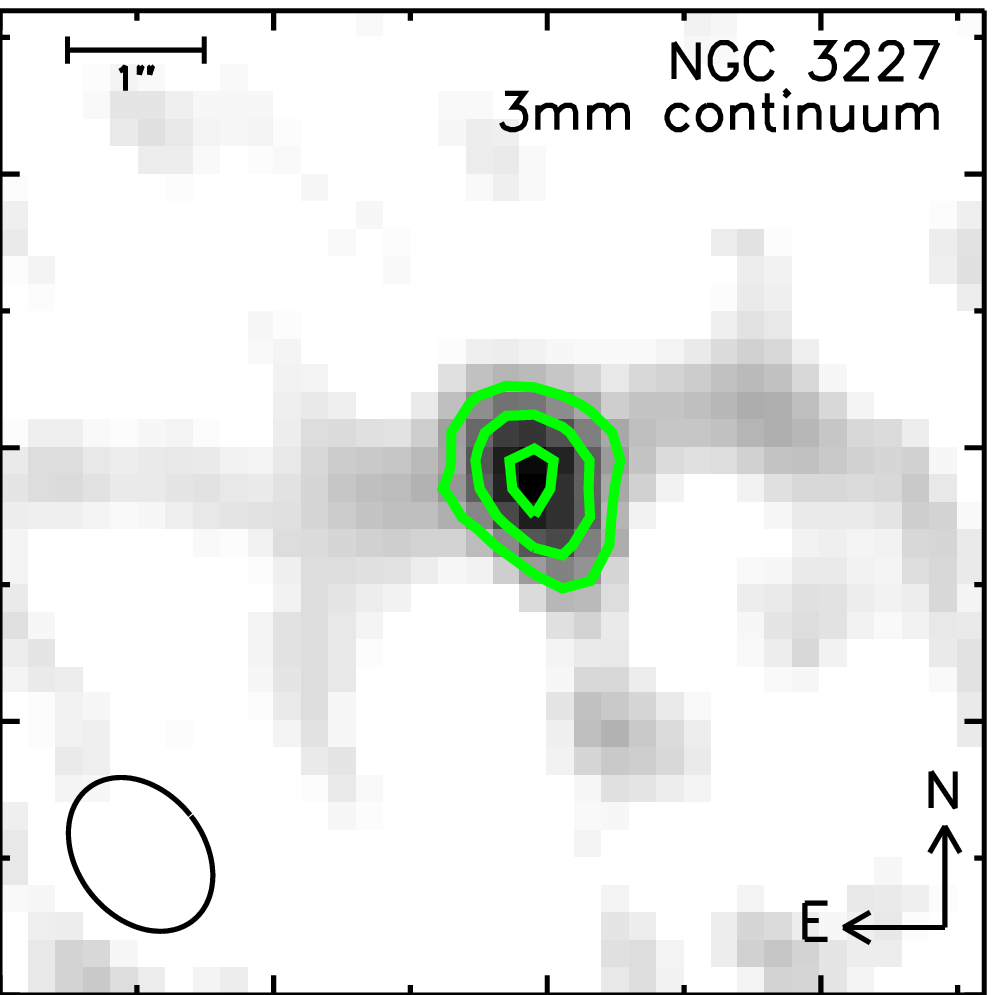}
\includegraphics[height=0.3\linewidth]{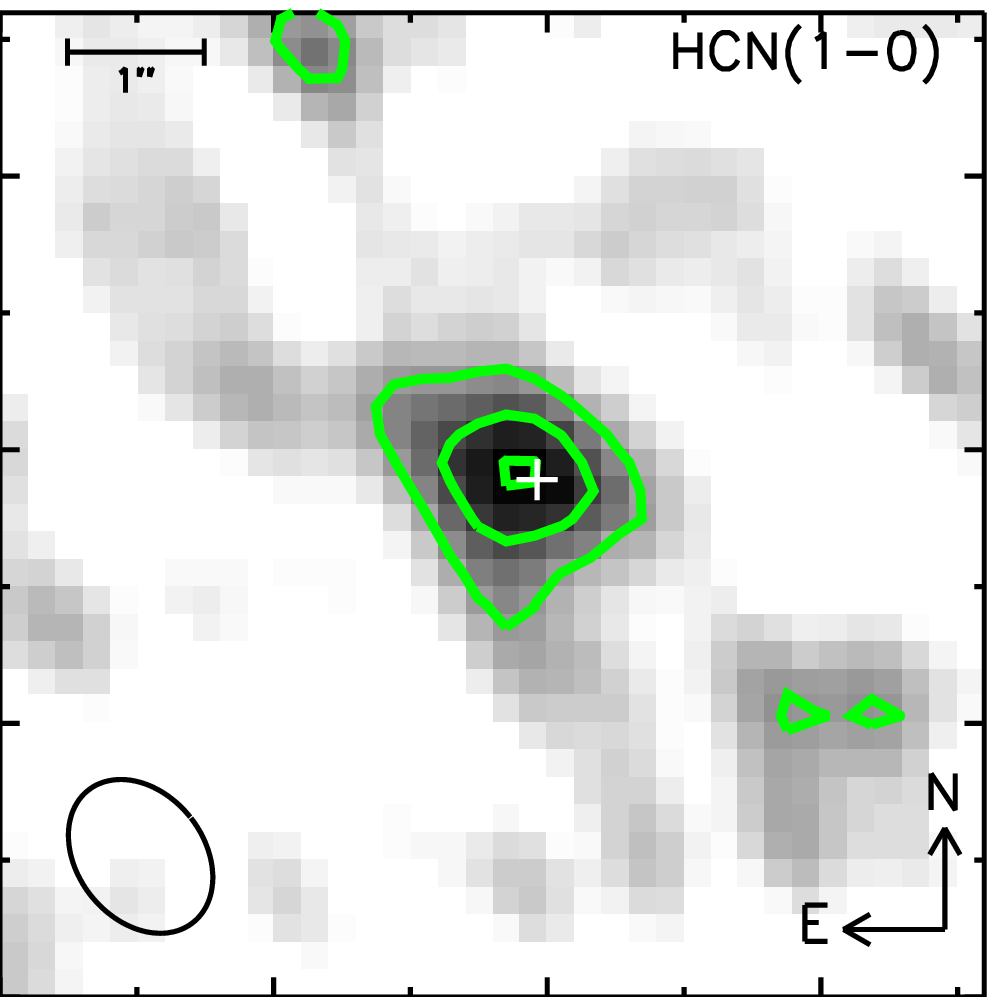}
}
}
\caption{The 3~mm continuum (left, rms~$=0.07$~Jy~beam$^{-1}$km~s$^{-1}$) and \hcn\ 
(right, rms~$=0.11$~Jy~beam$^{-1}$km~s$^{-1}$) emission maps 
of NGC~3227. Labels are as in Fig.~\ref{fg:2273}. 
The contour levels in the left panel are at 3, 5, and 7 times the noise level; in the right panel at 2, 4, and 6 times the noise level.
The white `plus' sign in the right panel indicates the location of the continuum peak. 
The continuum  is unresolved while the line emission is clearly extended.}
\label{fg:3227}
\end{figure*}

The 3~mm continuum appears to be partially resolved, having a FWHM of $1.86\arcsec\times1.13\arcsec$ at a PA$=87^\circ$ (from a Gaussian fit 
to the data in the left panel of Fig.~\ref{fg:2273}).
The 3~mm flux density of $2.46\pm0.16$~mJy from a Gaussian fit to the continuum map is consistent with 
the mean spectral value of 
$2.34\pm0.09$~mJy from the integrated spectrum (Fig.~\ref{fg:sp2273}). 

The \hcn\ line (Fig.~\ref{fg:2273}, middle panel) has an integrated flux of $1.76\pm0.07$~Jy~km/s 
in a $3\arcsec$ aperture. Its FWHM is $1.87\arcsec\times 1.17\arcsec$ at a position angle of $18^\circ$.
Allowing for the beam size, this is consistent with the projected intrinsic size of $1.3\arcsec\pm0.3\arcsec$ from a symmetric Gaussian fit to the $uv$ points.
Thus the line emitting region is compact but nevertheless spatially resolved. 

This conclusion is confirmed by comparing the \hcn\ properties with those of the \hco\ 
line (Fig.~\ref{fg:2273}, right panel), which has an integrated flux of $2.13\pm0.08$~Jy~km/s in a $3\arcsec$ aperture.
The latter has a FWHM of $1.94\arcsec\times 1.34\arcsec$ that, 
given the errors derived with Monte Carlo realizations listed in Table~\ref{tb:lines}, is consistent 
with the corresponding measurement above for the \hcn\ line with respect to the major axes, but implies a more extended source along the minor axis. 
A symmetric Gaussian fitted to the $uv$ table gives an intrinsic FWHM of $1.7\arcsec\pm0.4\arcsec$, also in agreement with the size based on spatial coordinates. 

The position angle of the \hco\ major axis appears to differ slightly from that of the \hcn,  and their centers are marginally offset with respect to the continuum.
This could reflect different distributions of the tracers even though the offset observed is of the same order of the positional accuracy. 
On the other hand, the PAs of the red/blue channels for the two lines (see Sec.~\ref{sec:kinematics}) are in much better agreement. 
The \hcn\ and \hco\ lines are also similar spectrally. 
Indeed a single Gaussian fit to the spectra in Fig.~\ref{fg:sp2273} gives FWHMs of 181~km/s 
and 175~km/s for \hcn\ (top panel) and \hco\ (bottom panel) respectively. 
We have derived uncertainties of 21~km/s and 16~km/s with Monte Carlo techniques.

\subsection{NGC~3227}

\begin{figure}
\hbox{
\includegraphics[height=0.6\linewidth]{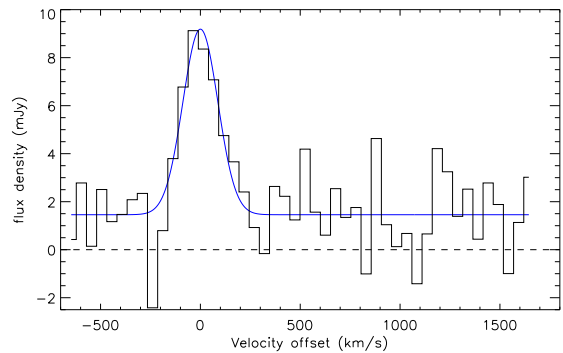}
}
\caption{Integrated spectrum of NGC~3227 showing the 3~mm continuum and the \hcn\ 
line with a Gaussian fit (in blue) showing the large velocity dispersion of the line (FWHM=207~km/s).}
\label{fg:sp3227}
\end{figure}

\begin{figure*}
\centerline{
\hbox{
\includegraphics[scale=1]{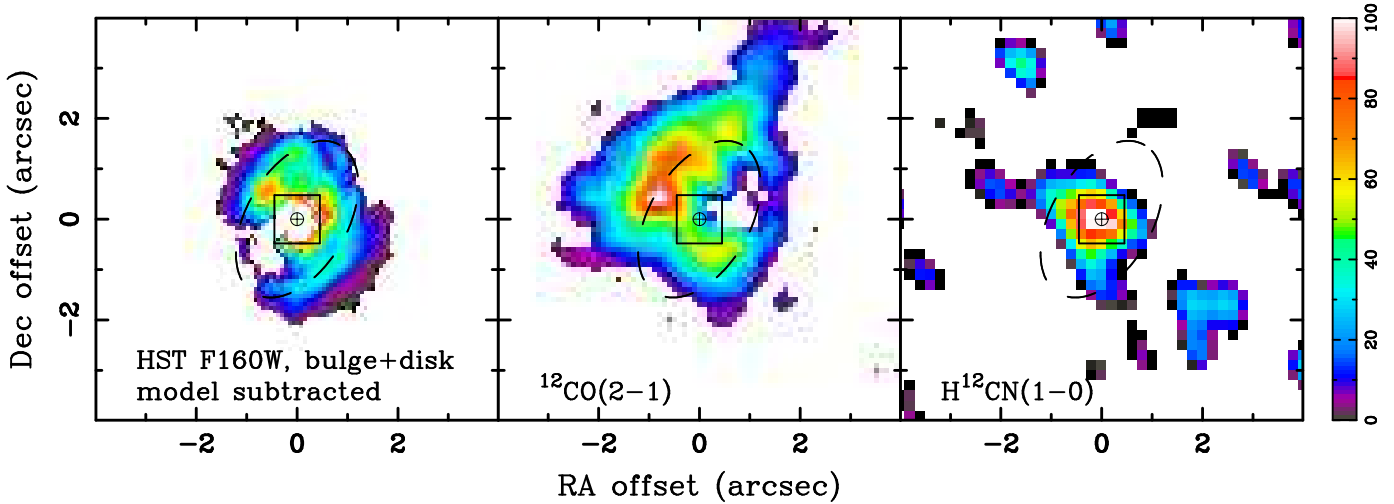}
}
}
\caption{Maps of the central few arcsecs of NGC~3227 ($1\arcsec\sim 80$~pc), adapted from \citet{dav06}.
In each panel the dashed ellipse traces the circumnuclear ring, 
and the square indicates the field of view of the SINFONI data presented by these authors. 
Left: H-band image of the circumnuclear ring (obtained after subtracting a bulge and disk 
model fitted to larger scales). Center: the CO(2-1) emission (data from \citealt{sch00a})  
is distributed around the ring, with relatively little from the nucleus. Right: the \hcn(1-0) 
peaks at the nucleus with relatively little originating in the ring. North is up and East 
is to the left (i.e., negative offsets).}
\label{fg:maps}
\end{figure*}

\begin{figure}
\hbox{
\includegraphics[height=0.6\linewidth]{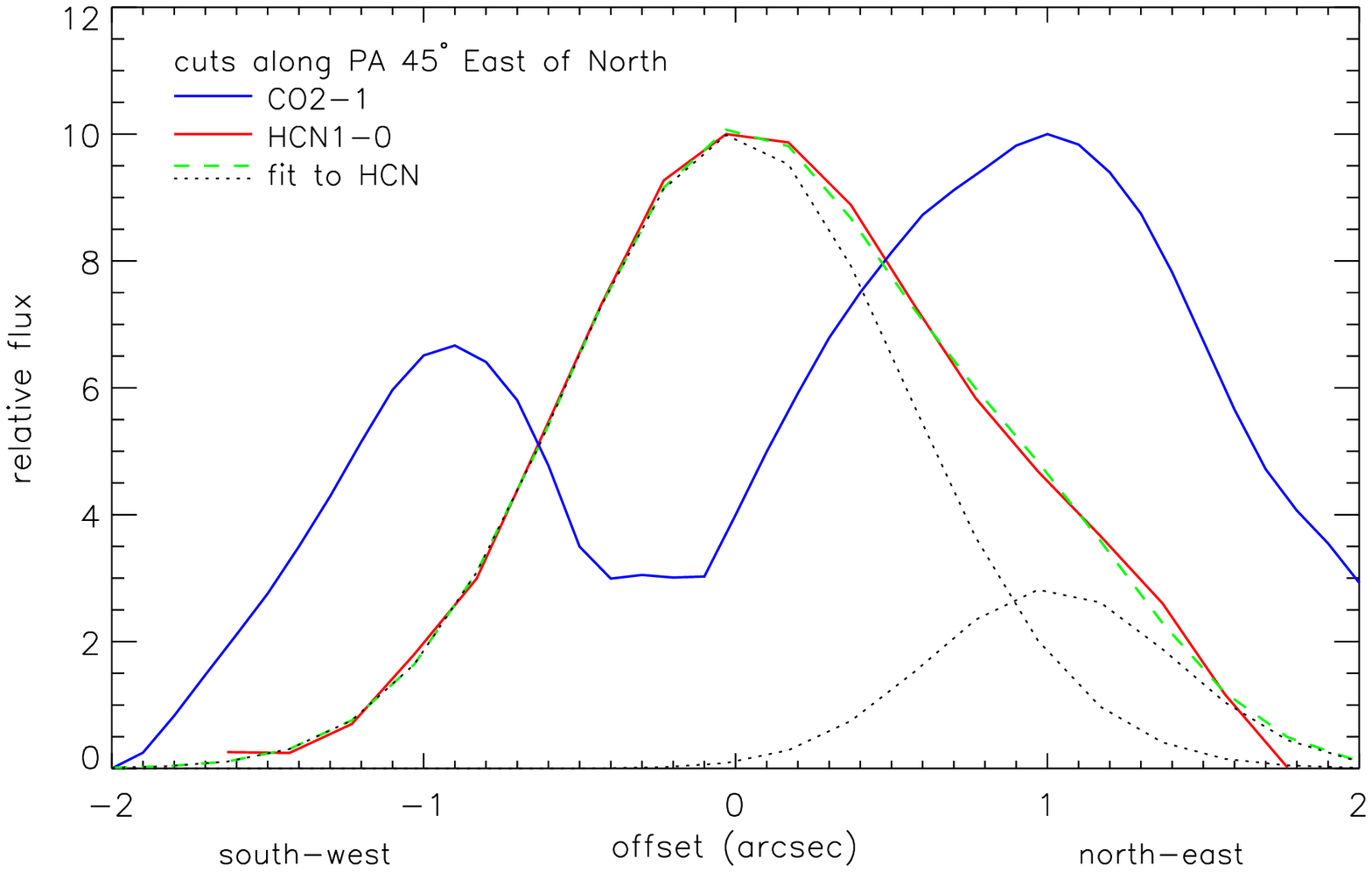}
}
\caption{NGC~3227 profiles of the CO(2-1) (solid blue line) and the \hcn(1-0) (solid red line) 
extracted in a $0.6\arcsec$ wide slit along a position angle of $45^\circ$ 
(i.e. along the minor axis of the circumnuclear ring). 
The CO peaks in the ring, while the \hcn\ peaks on the nucleus. 
Superimposed is a fit to the \hcn\ profile (dashed green line) which comprises 
3 Gaussians (dotted black lines; one of the Gaussians is not seen because it has zero flux). 
The secondary peak on the right, associated with the ring, matches the broad asymmetric overall profile. 
Taking just the central Gaussian yields an observable size for the 
nuclear emission of $1.28\arcsec$ at this PA.} 
\label{fg:prof}
\end{figure}

A Gaussian fit to the 3~mm continuum (Fig~\ref{fg:3227}, left panel) yields a size 
of $1.15\arcsec\times 1.05\arcsec$ FWHM at a position angle of $47^\circ$. 
This is very similar to the beam size and indicates the continuum is spatially unresolved. 
The 3~mm flux density 
from the integrated spectrum (Fig~\ref{fg:sp3227}) is $1.56\pm0.24$~mJy, consistent 
with the $1.79\pm0.12$~mJy obtained from the Gaussian fit to the continuum map.
 
The \hcn\ line has an integrated flux of $1.86\pm0.27$~Jy~km/s in a $3\arcsec$ aperture 
(Fig~\ref{fg:3227}, right panel). Comparison with the previous measurement of 2.1~Jy~km/s 
at $2.4\arcsec$ resolution \citep{sch00a} 
suggests that very little of the line emission has been resolved out at our higher resolution, and that most of the 
\hcn\ in NGC~3227 does originate from the central compact source. 
The compact nature of the \hcn\ emission is in stark contrast to the CO(2-1) emission, 
as shown in Fig.~\ref{fg:maps}. Indeed the CO emission is distributed around the circumnuclear 
ring (which is also seen in the H-band stellar continuum, Fig~5 and \citealt{dav06}), 
and very little originates from the central arcsecond. On the other hand, 
the \hcn\ emission is dominated by the nucleus itself.

A symmetric Gaussian fit to the data in the $uv$ plane yields a projected 
intrinsic FWHM of $0.8\arcsec\pm0.3\arcsec$. 
This is consistent with the FWHM of $1.5\arcsec\times 1.04\arcsec$ (at PA $47^\circ$) measured from the reconstructed image, once the finite beam size is taken into account.
It indicates that the line emission is resolved. 
However, we note that the long axis of the nuclear emitting region coincides with the minor axis 
of the circumnuclear ring, which as Fig.~\ref{fg:maps} shows is traced by the CO(2-1) emission. 
Along this axis, \hcn\ emission from the ring may be blended with that from the nucleus, 
which would bias the size measurement of the nuclear source. We have therefore separated these contributions 
by extracting a profile along this position angle (see Fig.\ref{fg:prof}) and fitting 3 independent Gaussians at fixed positions -- representing the nucleus and a cut through the ring on either side -- to the overall \hcn\ profile. 
One of these, which would be associated with the ring to the south-west, 
has a negligible contribution and so does not appear in the plot. 
However, the north-east side of the \hcn\ profile is clearly broadened by a subsidiary peak.
Since the asymmetry of the full profile can be matched by the addition of a component at the same location as the CO (which arises from the ring), we conclude that it is associated with the ring. 
Accounting for this reduces the observed FWHM of the nuclear component slightly to 
$1.28\arcsec$. Quadrature correcting the observed $1.28\arcsec\times 1.04\arcsec$ size of this component for the beam 
indicates that the intrinsic source size may be as small as $0.5\arcsec$ along both axes, a little 
less than implied by direct fitting of the $uv$ data. 

As found in NGC~2273, the \hcn\ line in NGC~3227 has a remarkably large velocity width. 
A simple Gaussian fit to the integrated spectrum in Fig.~\ref{fg:sp3227} gives 
a FWHM of $207\pm34$~km/s, where, as before, we have used Monte Carlo techniques to estimate the uncertainty.

\subsection{NGC~4051}

\begin{figure*}
\centerline{
\hbox{
\includegraphics[width=0.3\linewidth,angle=0]{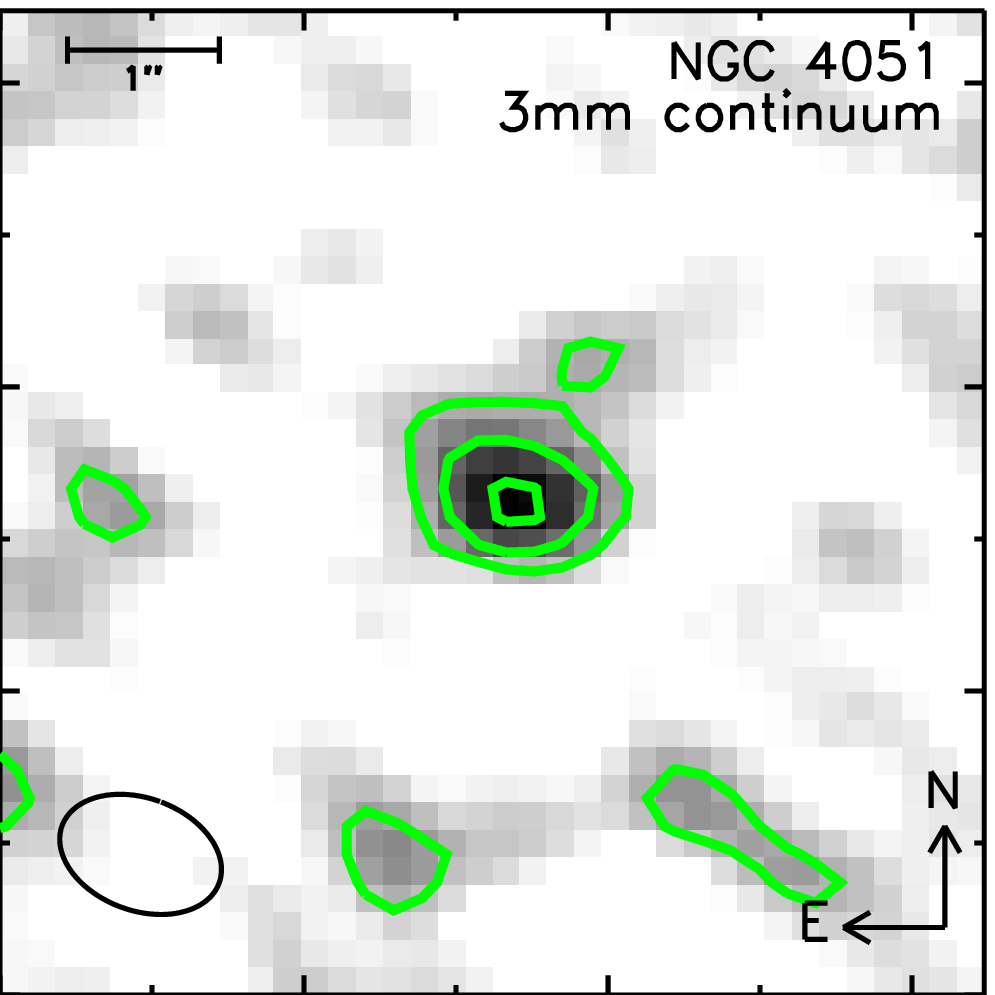}
\includegraphics[height=0.3\linewidth]{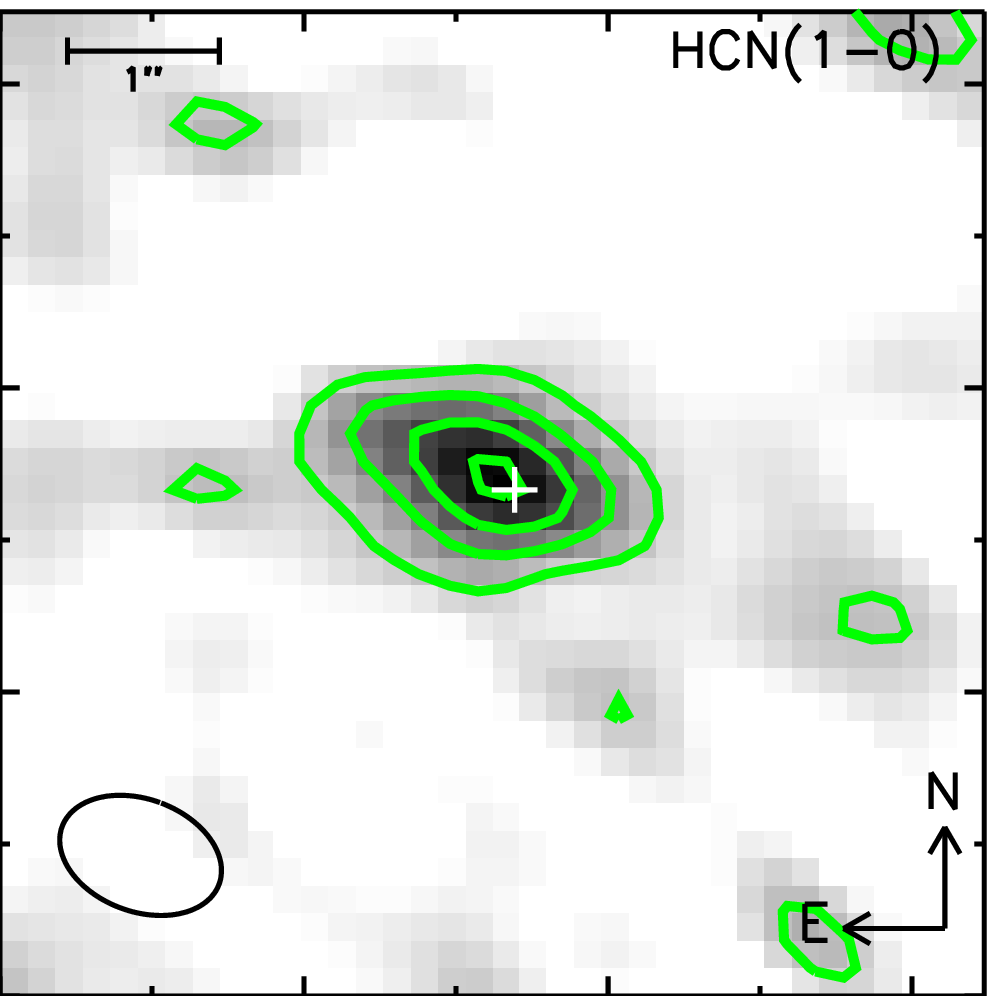}
\includegraphics[height=0.3\linewidth]{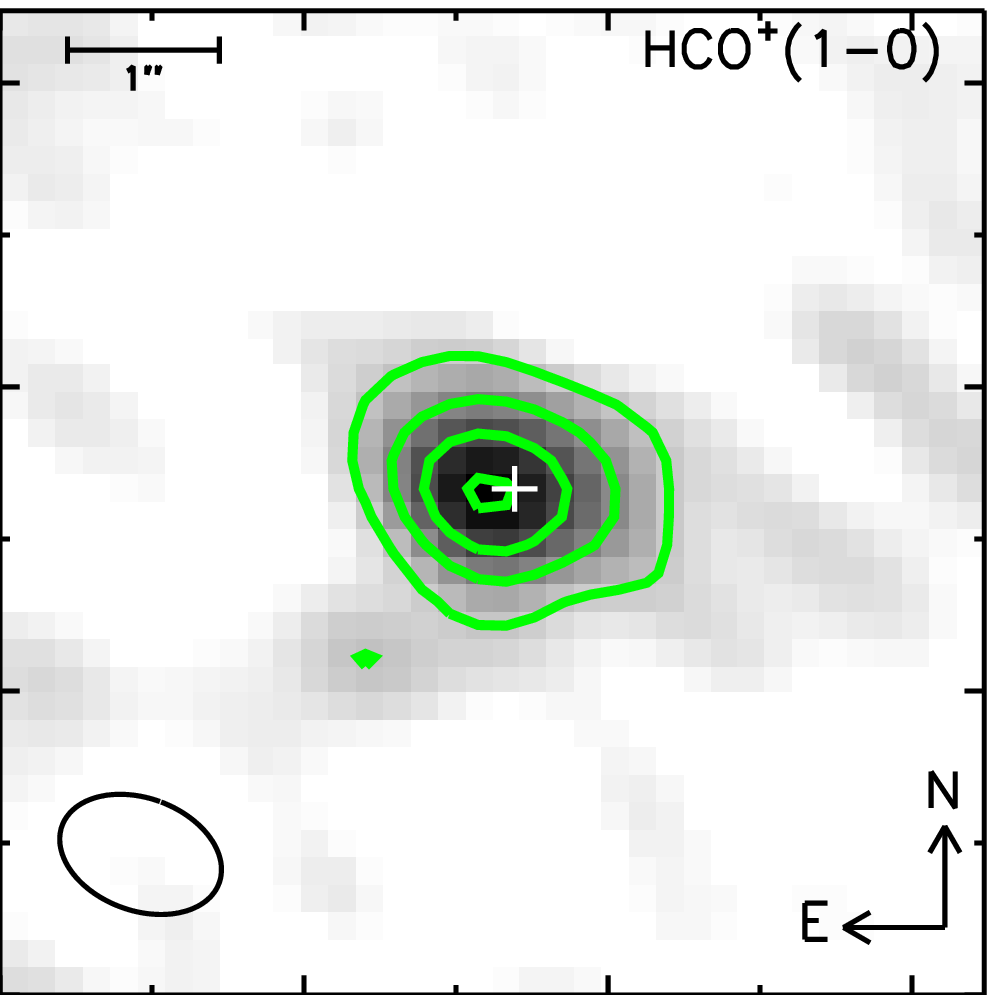}
}
}
\caption{The 3~mm continuum (left, rms~$=0.02$~Jy~beam$^{-1}$km~s$^{-1}$), 
\hcn\ (middle, rms~$=0.04$~Jy~beam$^{-1}$km~s$^{-1}$) and 
\hco\ (right, rms~$=0.05$~Jy~beam$^{-1}$km~s$^{-1}$) emission maps 
of NGC~4051. 
Labels are as in Fig.~\ref{fg:2273}. 
The contour levels run between 2--6 times (left) and 2--8 times (middle) the noise level
in steps of $2~\sigma$;
and between 3--12 times (right) the noise level in steps of $3~\sigma$.
The white `plus' signs denote the location of the continuum peak.
The continuum is basically unresolved while the line emission is clearly extended.}
\label{fg:4051}
\end{figure*}

\begin{figure}
\vbox{
\includegraphics[height=0.6\linewidth]{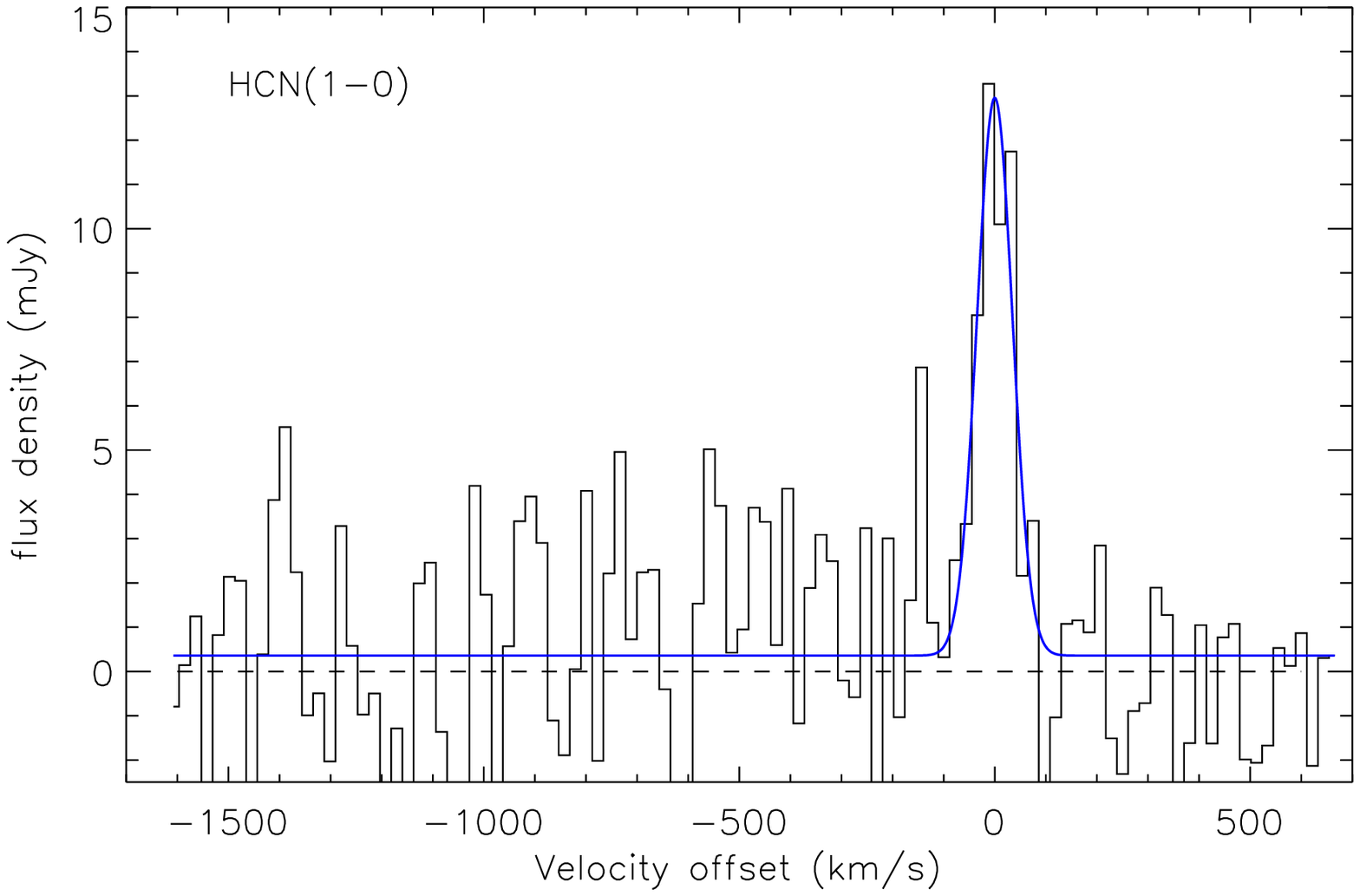}
\includegraphics[height=0.6\linewidth]{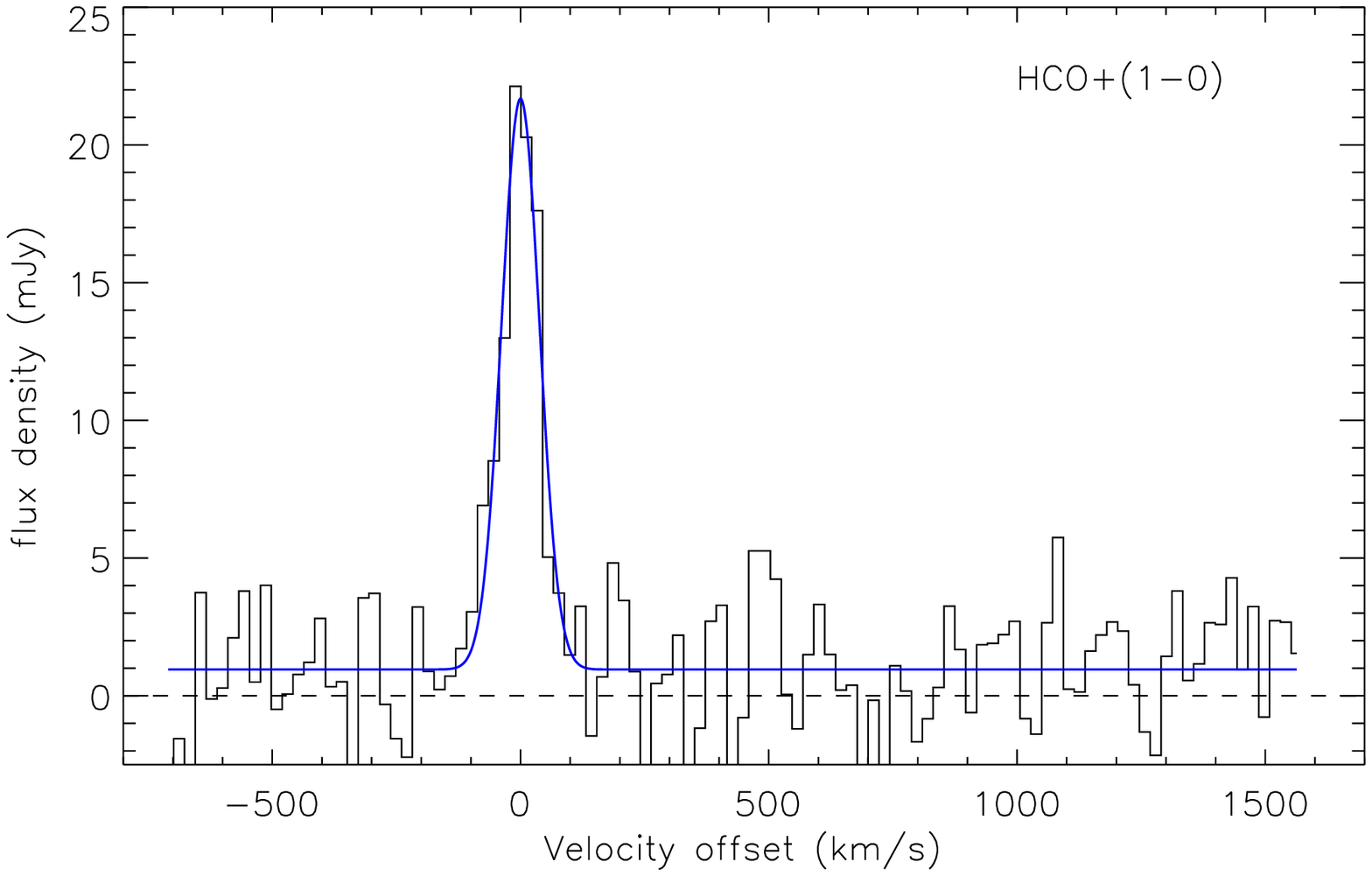}
}
\caption{\hcn\ (upper) and \hco\ (lower) integrated spectra of NGC~4051. 
Compared to NGC~2273 (Fig~\ref{fg:sp2273}) and NGC~3227 (Fig~\ref{fg:sp3227}) both 
the continuum level and mean line width of FWHM=$82\pm10$~km/s are a factor of 2 lower. 
Lines are as in Fig~\ref{fg:sp2273}.}
\label{fg:sp4051}
\end{figure}

A Gaussian fit to the continuum map of NGC~4051 (Fig.~\ref{fg:4051}, left panel), 
gives a FWHM of $1.13\arcsec\times0.90\arcsec$ at PA$=82^\circ$.
Comparing this to the $1.07\arcsec\times0.72\arcsec$ beam 
at $80^\circ$ suggests the 3~mm continuum is basically unresolved. 
The mean flux density of $0.9\pm0.2$~mJy measured from the integrated spectra (Fig.~\ref{fg:sp4051}) is consistent with the $1.02\pm0.07$~mJy from 
a Gaussian fit to continuum map. 

Symmetric Gaussians directly fitted to the $uv$ data give FWHMs of $0.85\arcsec$ and $1.35\arcsec$ for the projected intrinsic size of the \hcn\ and \hco\ lines respectively 
(we note that, as before, because these sizes are derived from the {\it uv} data, they are free of beam convolution).
The molecular lines trace almost the same region as testified by the fits in spatial 
coordinates (Fig~\ref{fg:4051}, central and right panels):
the \hcn\ emission has an extent of $1.92\arcsec\times1.02\arcsec$ at PA=$80^\circ$, 
similar to $1.65\arcsec\times1.27\arcsec$ at PA=$72^\circ$ for the \hco\ emission. 
Both the \hcn\ 
and \hco\ emission are clearly resolved when compared to the beam size. 
The velocity dispersion, obtained by fitting the spectra 
in Fig.~\ref{fg:sp4051}, is about half that of either NGC~2273 or NGC~3227. 
The FWHM is indeed $74\pm10$~km/s and $90\pm10$~km/s for the \hcn\ and \hco\ lines respectively.

In NGC~4051 the line fluxes are $0.91\pm0.05$~Jy~km/s for \hcn, and $2.07\pm0.05$~Jy~km/s for \hco, both integrated over the central $3\arcsec$.
This means that the \hcn/\hco\ intensity ratio in NGC~4051 differs significantly from NGC~2273 where the ratio is close to 1.
This aspect of the line emission will be assessed, together with ratios of other line transitions, in a future paper.

\subsection{NGC~6951}

\begin{figure}
\centerline{
\hbox{
\includegraphics[height=0.65\linewidth]{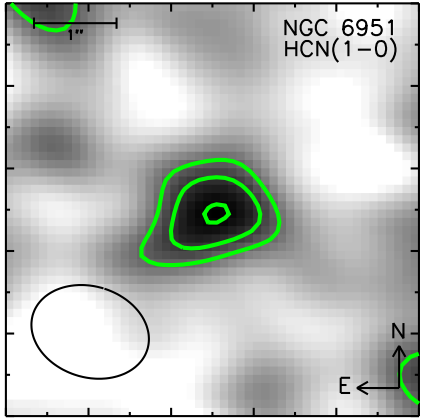}
}
}
\caption{The \hcn\ emission map (rms~$=0.07$~Jy~beam$^{-1}$km~s$^{-1}$)
of NGC~6951. Labels are as in Fig.~\ref{fg:2273}. 
Contour levels are at 3, 4, and 5 times the noise level.
The line emission appears at best only marginally resolved.}
\label{fg:6951}
\end{figure}

\begin{figure}
\hbox{
\includegraphics[height=0.6\linewidth]{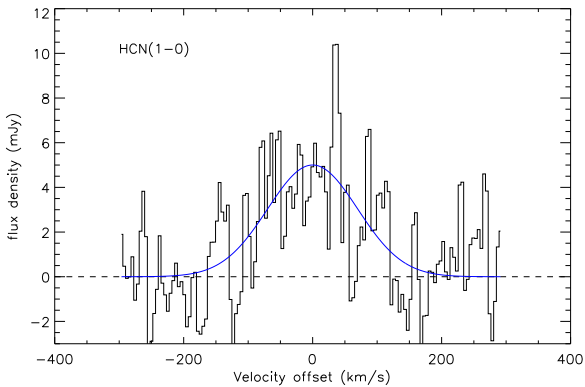}
}
\caption{Integrated spectrum of NGC~6951 showing the \hcn\ 
line (a fit for which is overdrawn in blue), with its large velocity dispersion (FWHM=173~km/s).}
\label{fg:sp6951}
\end{figure}

The properties of this source have already been presented by K07. 
As stated in that work, reconstructing an image with uniformly weighted visibilities provides high enough resolution that the emission from the ring does not contaminate the nuclear signal.
Here we briefly summarize the main characteristics of the nuclear emission that are required for our modelling, noting that the values listed in Table~\ref{tb:lines} agree very well with the K07 analysis and references therein. 
The \hcn\ central emission (Fig.~\ref{fg:6951}) originates from a region of 
$1.44\arcsec\times1.10\arcsec$ at PA = $73^\circ$, at best only marginally resolved with respect to the beam (Table~1), 
with an integrated flux of $1.02\pm0.02$~Jy~km/s  within $3\arcsec$.  
A single Gaussian is used to fit the line profile (Fig.~\ref{fg:sp6951}), and yields a FWHM of $173\pm29$~km/s. 

\section{Physical properties of the dense gas}
\label{sec:kinematics}

\begin{figure}
\hbox{
\includegraphics[height=0.5\linewidth]{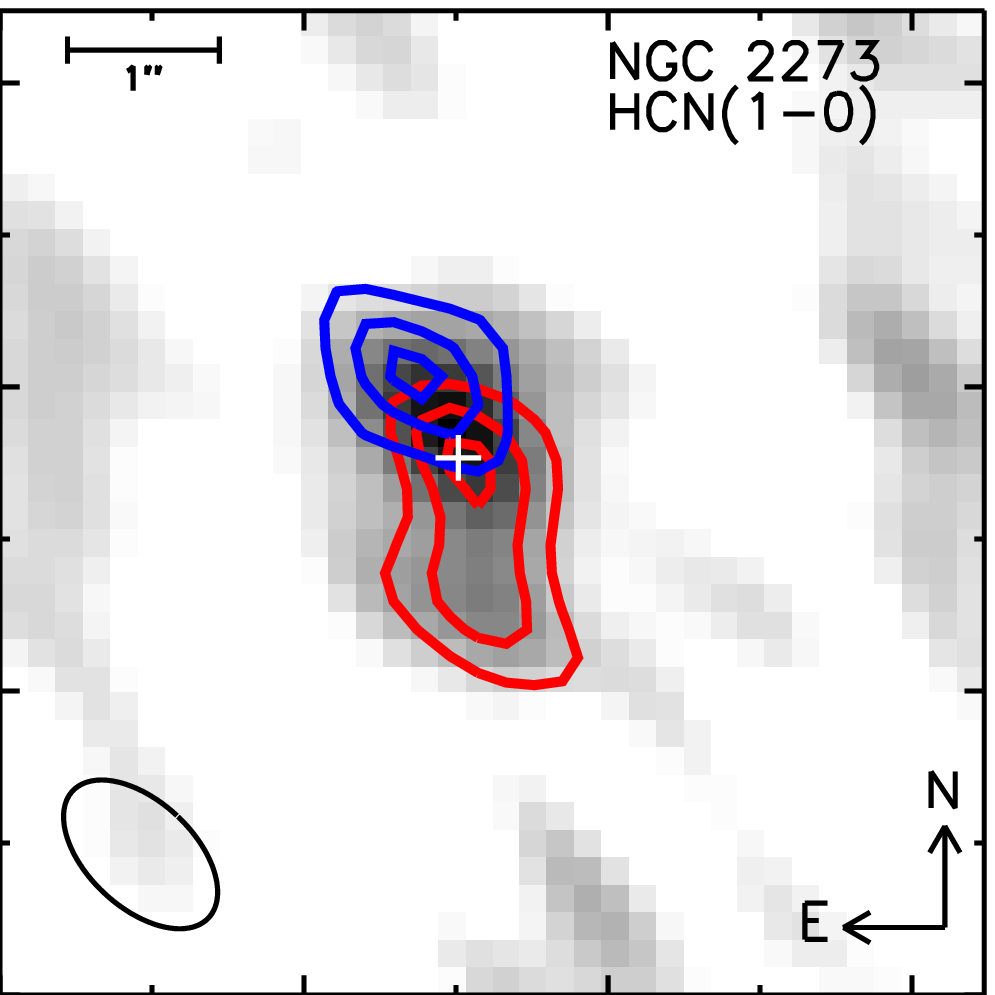}
\includegraphics[height=0.5\linewidth]{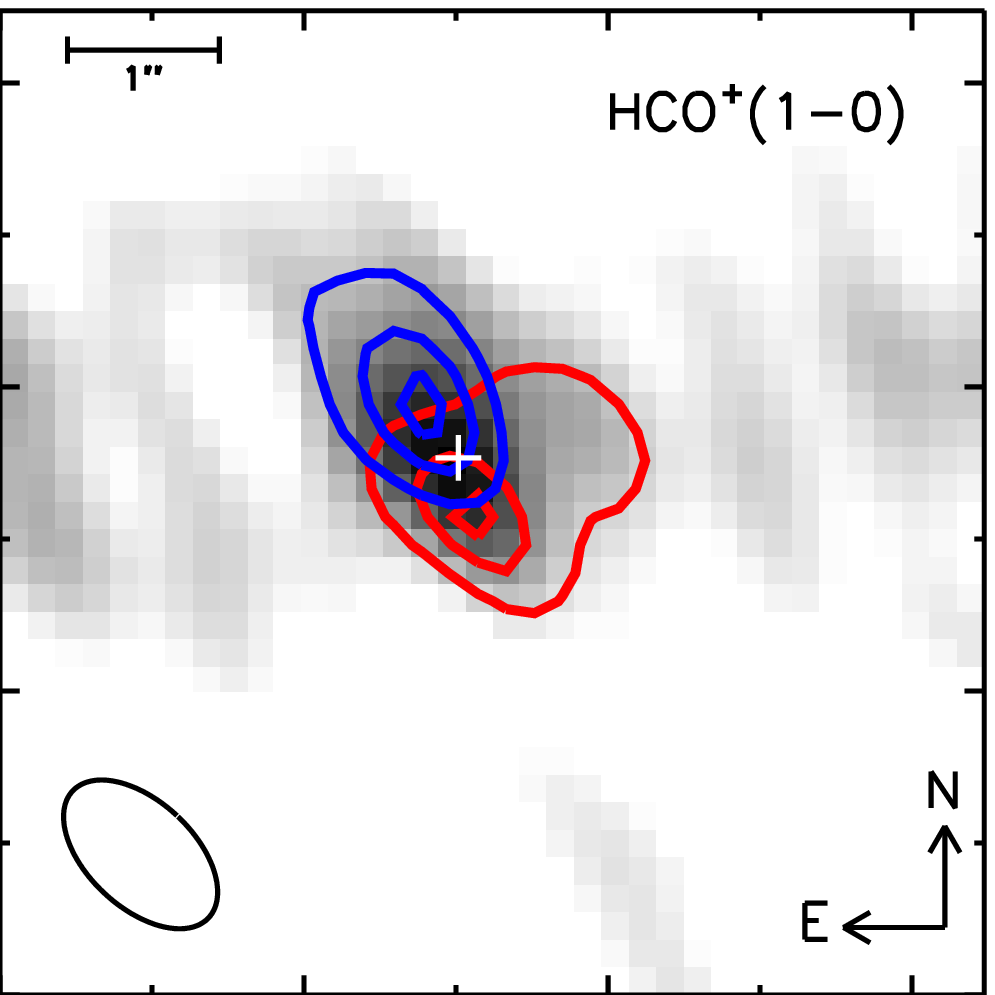}
}
\hbox{
\includegraphics[height=0.5\linewidth]{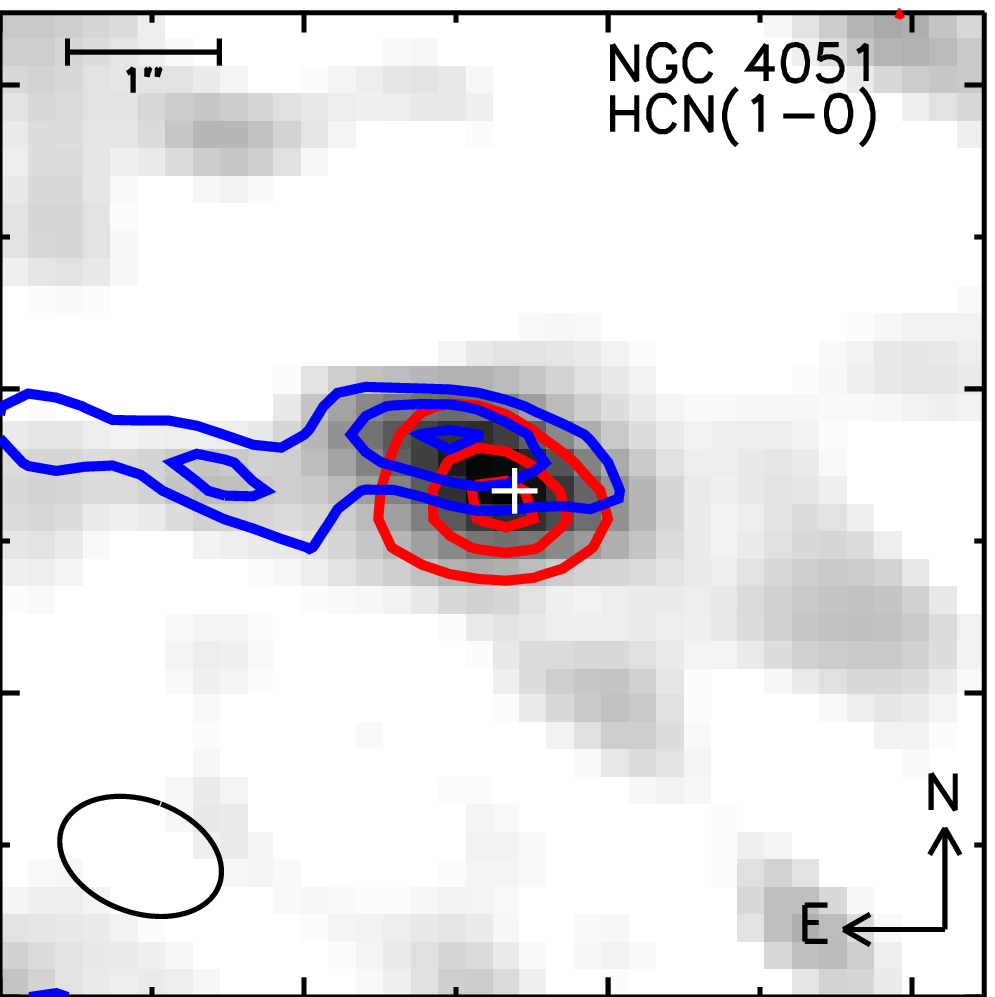}
\includegraphics[height=0.5\linewidth]{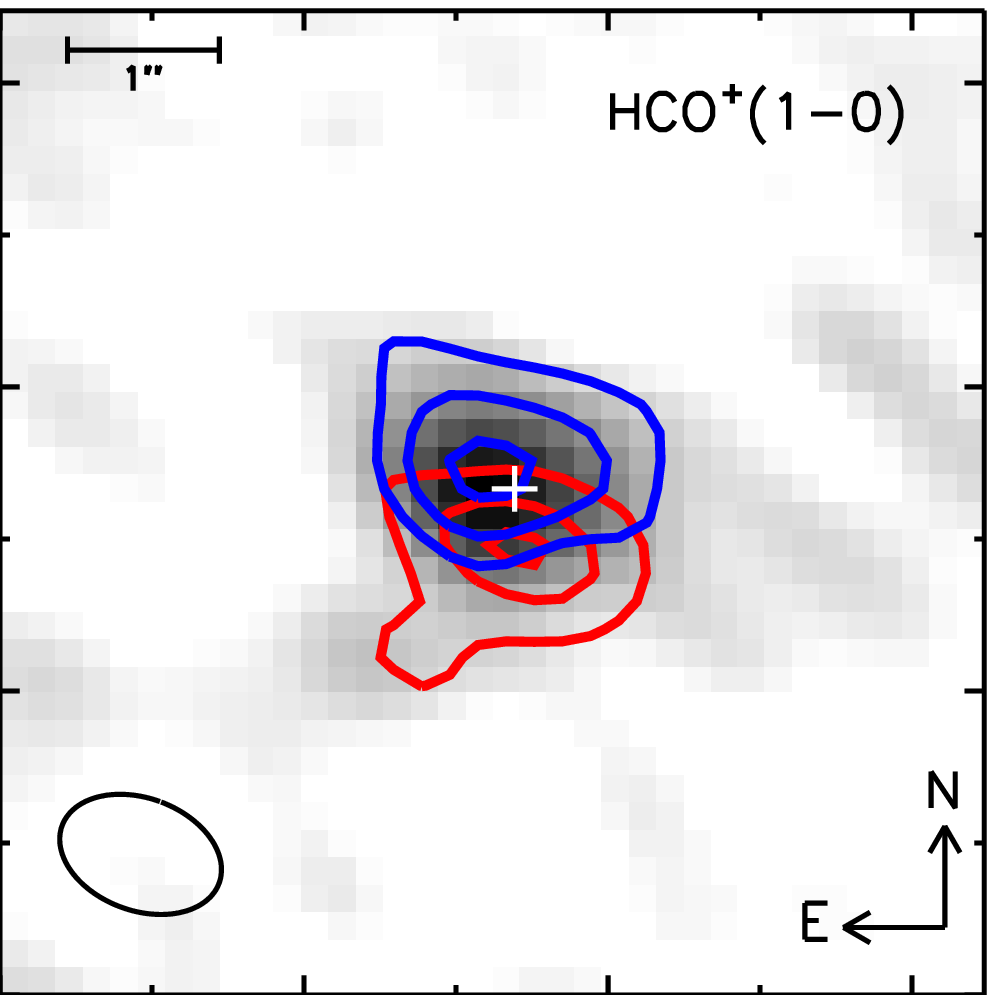}
}
\hbox{
\includegraphics[height=0.5\linewidth]{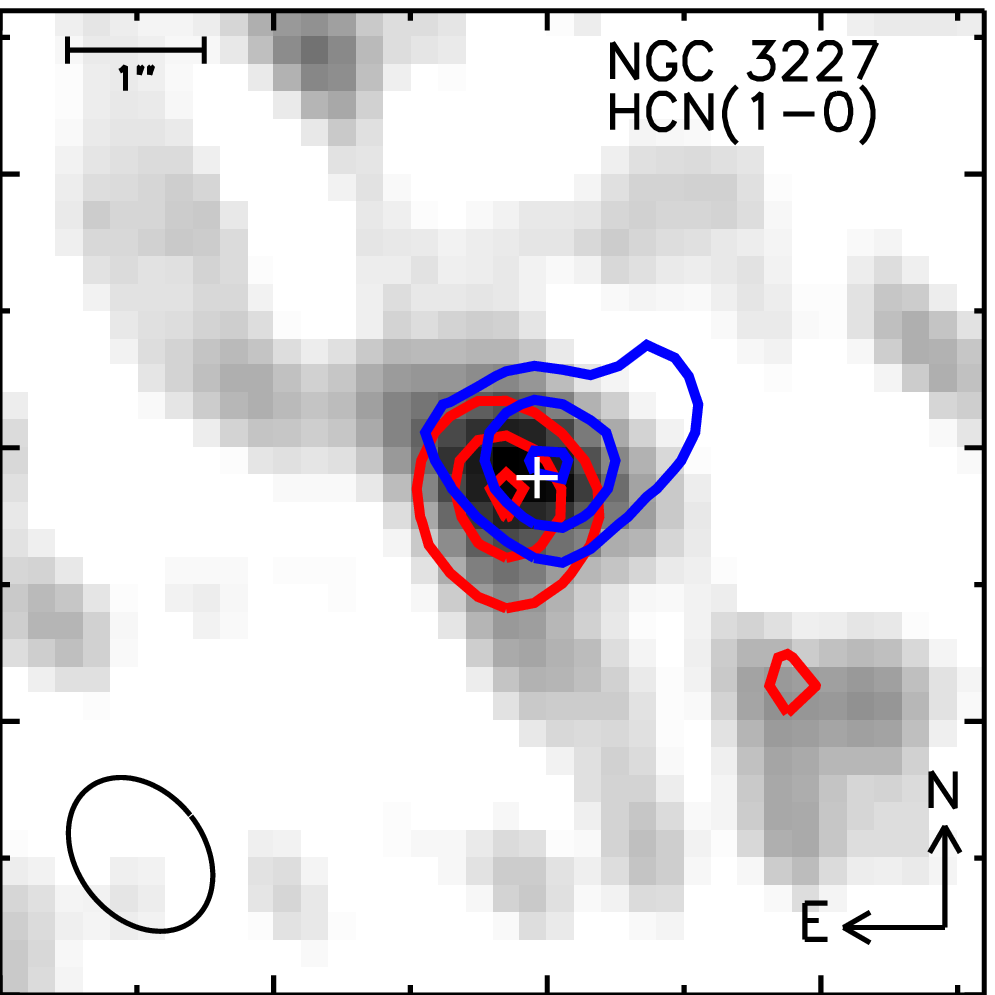}
\includegraphics[height=0.5\linewidth]{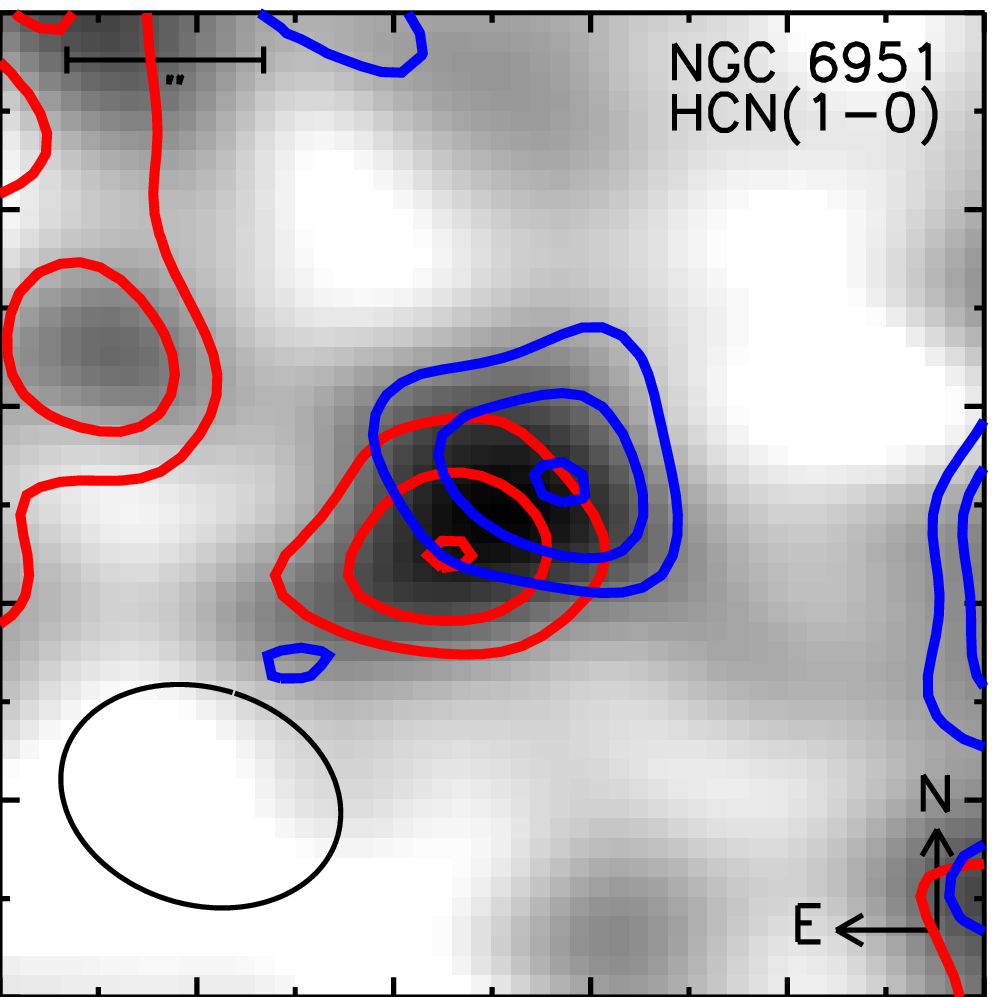}
}
\caption{Channel maps of \hcn\ and \hco, superimposed on a 
greyscale image of the line. From top to bottom: NGC~2273, NGC~4051, NGC~3227 (left) and NGC~6951 (right). 
The red and blue channels 
are summed over different ranges from source to source, based on the line profile: $\pm(20-90)$~km/s (\hcn) $\pm(20-100)$~km/s (\hco) for NGC~2273, 
$\pm(50-200)$~km/s for NGC~3227, 
$\pm(20-70)$~km/s (\hcn) $\pm(20-80)$~km/s (\hco) for NGC~4051, and $\pm(10-200)$~km/s for NGC~6951 (where the different velocity resolution of 
the data implies a different channel binning). 
The contour levels are at $2~\sigma$, $~3\sigma$, and $5~\sigma$ for NGC~2273, NGC~3227 and NGC~6951, 
while for NGC~4051 are at $~3\sigma$, $5~\sigma$ and $~7\sigma$.
The white `plus' signs denote the peak of the continuum.
In all cases the peaks of the red and blue channel maps are well separated, showing that the velocity gradient is resolved.}
\label{fg:gr}
\end{figure}

In this Section we first compare the dense gas kinematics to the 
warm gas and stellar kinematics. We then describe the dynamical models we apply to the data, 
under the assumption that the gas is distributed in a rotating disk.

The high spectral resolution of the data enables us to measure a velocity gradient 
even though the sources are only marginally resolved, by generating red and blue channel 
maps -- a technique that is equivalent to the spectroastrometry used at optical wavelengths.
It is a well-known technique that has been applied in many different circumstances, and is able to trace 
velocity gradients on spatial scales much smaller than the spatial beam or seeing.
The gain can easily be a factor 10 or more (e.g. see \citealt{gne10}), since the measurement simply 
depends on the precision with which the centroid 
of the emission can be measured, which is typically possible to $\sim1/10$ of the beam size. 
The channel maps are generated by summing the line flux over positive and negative velocity ranges with respect to the line center (zero velocity).    
Figure~\ref{fg:gr} shows the channel maps of both \hcn\ and \hco\ for all targets and Table~\ref{tb:lines} lists the relative centroids separation 
and position angle of the vector joining them. 
We note that the centers of red and blue channel maps are derived each time from the same dataset, and are spectrally close to each other. 
Therefore any offsets between red/blue emission due to phase calibration are negligible and 
the uncertainties on their separation and relative position angle are only due to centroid positioning. 
For each object and emission line, we derive the errors on centroid position running Monte Carlo realizations of the 
red and blue channel maps, as described in Section 3. We then estimate uncertainties on their separation and relative position angle 
using a standard error propagation. 
In all the cases, the red and blue centroids are well separated, in most cases by a significant fraction of the beam sizes given in Table~1. 
The velocity gradients are thus well resolved.
However, we note that the red and blue maps are not always centered symmetrically about the continuum, the location of which is indicated by 
the white plus signs on the maps (excluding NGC~6951, for which we have no continuum measurement).
This is related to the apparent off-centre nature of the line maps in earlier figures, which may indicate that the \hcn1-0 and \hco1-0 emission is not exactly axisymmetric, as might be expected if there are spatial variations in gas density, temperature or excitation.

For NGC~3227 the separation of the centres of the red- 
and blue-shifted emission is $0.38\arcsec$, and the velocity gradient is oriented similarly 
to that of the stars \citep[hereafter B06]{dav06,dav07,bar06}
and warm gas as traced by the H$_2$~(1-0)\,S(1) line (H09). 
Similarly, the centroid separation of $0.55\arcsec-0.60\arcsec$ in NGC~4051 is consistent with the stellar kinematics (B06), 
but there is a mismatch in its position angle with respect to the warm gas and 
stellar kinematics in H09. 
This discrepancy can be understood in terms of both the angle measurement error ($\pm10^\circ$), 
and the field of view of the IFU ($0.56\arcsec\times2.24\arcsec$, H09) compared to the interferometer beam size in Table~1. 
For NGC~2273 we can compare the dense gas kinematics only with the stellar component 
(B06), and these are consistent. 
For NGC~6951 stellar and warm gas kinematics are not available, and so we compare the \hcn\ kinematics to those of the ionized gas reported by \cite{sto07}.
The peaks of the red and blue \hcn\ channel maps are separated by $0.60\arcsec$ at a PA 
consistent with the line of nodes inferred for the ionized gas. 
We can thus conclude that in our nuclei, the velocity gradients of the dense gas are, as far as can be ascertained, consistent with those observed with other tracers.
 
\begin{figure}
\centerline{
\hbox{
\includegraphics[height=0.8\linewidth]{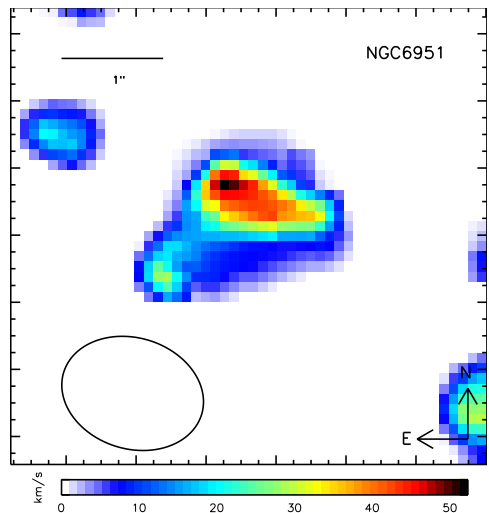}
}
}
\caption{\hcn\ velocity dispersion map of NGC~6951, which shows a ridge of high dispersion along the minor axis (the typical signature associated with severe beam smearing of a rotating disk), rather than along the radio jet (between the major and minor axes) as has been reported by \citet{sto07} for the outflowing ionised gas.}
\label{fg:sigma}
\end{figure}

The nuclear gas kinematics of our targets exhibit ordered rotation. 
For example, high resolution observations of NGC~3227 and NGC~4051 show that the PA of the warm H$_2$ 
remains constant throughout the nuclear region (H09).
This suggests that the nuclear gas has no significant radial motions (e.g., in/out flows) 
and no warp (i.e. there is no twisting of the kinematics major axis PA and/or inclination angle). 
The same conclusion is drawn in \cite{bar09}, where the sample includes also NGC~2273: 
although the ionized gas shows signs of outflowing motion, the kinematics are rather dominated by circular rotation, 
coplanar with the stellar component. 
In NGC~6951, 
\cite{sto07} found evidence of an outflow due to the interaction of a radio jet \citep{sai02}
with the circumnuclear ionized gas. 
They measured a significantly enhanced velocity dispersion (increasing from $\sim80$\,km/s to as much as 150\,km/s) in two blobs of gas at PA $156^\circ$, an orientation close to that of the \hcn\ velocity gradient. 
To verify if this feature is present in our \hcn\ data, we show in Fig.~\ref{fg:sigma} 
the velocity dispersion map of NGC~6951. 
The observed dispersion pattern has a ridge of high values perpendicular to the velocity gradient, 
which is the typical signature of a rotating disk that is subject to severe beam smearing. 
As such, in NGC~6951, the radio jet appears to have negligible influence on the dense gas kinematics on these scales.  
We therefore conclude that there is no evidence for dense gas in radial motion in the sources considered here.
Instead we expect \hcn\ and \hco to trace disk rotation. 

\begin{table*}{Input and Output Parameters of Dynamical Models}
 \begin{center}
 \begin{tabular}{lcccccccccc}
 \hline
  Source   & \multicolumn{2}{c}{Fixed}& & \multicolumn{3}{c}{Input}& & \multicolumn{3}{c}{Output}\\
  \cline{2-3} \cline{5-7} \cline{9-11}
           & $\theta$   &    PA  &  &width   &   height  & M$_{scale}$& & Major x Minor axis &   r/b sep  & FWHM  \\
           &   deg      &  deg   &  & $\arcsec$ & $\arcsec$ & Arb. Uni. & &$\arcsec\times\arcsec$& $\arcsec$& km/s\\
 \hline
  NGC~2273 & 51         &  24   & &   1.5    &  0.43      &     2.9 &  &   1.91 x 1.28      &  0.99      & 180 \\     
 \hline
  NGC~3227 & 55         &  -37  & &   0.54   &  0.14      &   1.1   &  &  1.27 x 1.05       &  0.38      & 207\\
 \hline
  NGC~4051 & 50         &   29  & &   0.88   &  0.25      &   1.9   &  &   1.88 x 1.09      &  0.58      & 88 \\
 \hline
  NGC~6951 & 40        &   -50  & &   0.99   &  0.24      &   2.1  &   &   1.52 x 0.96     &  0.62       & 172 \\
 \hline
 \end{tabular}
  \label{tb:model}
 \caption{Dynamical simulations for a thick rotating disk. (1) Source name.
 The inclination of the molecular disk from H$_2$ 1-0\,S(1) or stellar kinematics (2), 
 and the PA (east of north) of the velocity gradient (3), are fixed parameters. 
 The input parameters left free to vary are the disk diameter (4), thickness (5), 
 and the scaling factor of the rotation curve (6).
 We adopt a Gaussian form for the radial mass distribution.
 The diameter and thickness are specified as FWHMs in, and perpendicular to, the disk plane. 
 The model outputs are: the 
 projected major and minor axes (7), the red and blue channel separation (8), and the line width (9). 
 These can be compared directly with the observed quantities in columns (4), (6), and (8) of Table~\ref{tb:lines}.
 For sources with two emission lines detected,  
the output values fall in between the observed quantities measured in Tab~\ref{tb:lines} for \hcn\ and \hco.}
 \end{center}
\end{table*}

One intriguing issue is the differences between $\sigma$ measurements from different tracers.
The velocity dispersion of the dense gas is systematically 
smaller than that of both the warm gas and stars. Once beam smearing is taken into account 
(see Sec~\ref{sec:models}), 
this effect is even more pronounced and implies that \hcn\ and \hco\ have intrinsic velocity dispersions about 1/2 that of the warm gas and/or stellar component. 
This is perhaps to be expected, because of the high densities and lower temperatures traced by 
the \hcn\ (\hco) molecules compared to the H$_2$ 1-0\,S(1) line. 
One might speculate that there could be a stratified 
structure with the denser gas at lower scale heights (i.e. lower velocity dispersion). 
Alternatively, since the H$_2$~(1-0)\,S(1) emission traces the edges of individual molecular clouds, if these are being ablated, then the line might preferentially trace outflows. 
Since the ablated gas would form filamentary structures, this would also tend to increase the velocity dispersion. 
Although there is no sign of this in H09, it still remains a possibility.
Despite the dispersion of the \hcn\ (\hco) being lower than that of other tracers, it still remains high.
In order to understand this, we first assess how much of the observed line width is due to beam smearing of a velocity gradient. 

In the following we describe how we model the data, and how the model is constrained by four observed parameters.
When data for two molecular lines 
are available, we can simply take the mean values of the quantities listed in Table~\ref{tb:lines}. 
Indeed, allowing for the various uncertainties (i.e. Monte Carlo realizations plus absolute position errors), 
all the quantities for the two lines in NGC~2273 and NGC~4051 are almost similar. 
Once we have modelled the kinematics, we discuss the implications 
in the context of the AGN-star formation connection.

\begin{table*}{Toy models for NGC~3227}
 \label{tb:toy}
 \begin{center}
 \begin{tabular}{lcccccccc}
 \hline
      &   \multicolumn{3}{c}{Input} & & \multicolumn{3}{c}{Output}\\
      \cline{2-4} \cline{6-8}
      & width   &   height  & M$_{scale}$&  &Major x Minor axis &   r/b sep  & FWHM  \\
      & $\arcsec$ & $\arcsec$ & Arb. uni.  & &$\arcsec\times\arcsec$& $\arcsec$& km/s\\
 model 1 & 0.5    & 0.01     & 0.5              & & 1.24 x 1.03          & 0.50   & 111 \\
 model 2 & 0.5    & 0.01     & 1.0              & & 1.24 x 1.03          & 0.44   & 191 \\
 model 3 & 0.5    & 0.10     & 0.5              & & 1.26 x 1.05          & 0.37   & 184 \\
 \hline     
 \end{tabular}
 \caption{Toy models for NGC~3227 using thin and thick rotating disks. 
 The inclination and position angle of the disk are $55^\circ$ and $-37^\circ$ respectively. 
 The input parameters, fixed in the three examples, are the disk diameter (2), thickness (3), 
 and the scaling factor of the rotation curve (4).
 The shape of the rotation curve is set by adopting a Gaussian 
 distribution for the mass profile. 
 The diameter and thickness are specified as FWHMs in, and perpendicular to, the disk plane.
 The model outputs are: the 
 projected major and minor axis (5), the separation of the centroids of the red and blue channel maps (6), and the molecular line width (7).}
 \end{center}
\end{table*}

\subsection{Modelling the kinematics}
\label{sec:models}

The aim here is to understand the basic intrinsic properties of the emitting gas, specifically 
the origin of the linewidth. 
The model we use is the IDL code DYSMAL which is described in Appendix~A of \citet{dav11} and 
has been applied in a variety of cases 
involving rotating disks \citep{dav04a,dav04b,dav09,gen08,cre09}.

DYSMAL is specifically designed to quantify the impact of the spectral and spatial 
beam smearing on an axisymmetric rotating disk, and thus allows us to infer the \textit{intrinsic} properties 
of the dense gas from the \textit{observed} ones. 
From a given set of input structural parameters, the code computes a data cube with two spatial and one velocity axis, which can be analysed in the same way as the original data. 
The initial setup includes one or multiple mass/light components, parametrized by azimuthally symmetric functions 
(e.g., Sersic or Gaussian profiles; rings, etc.), the total rotationally supported mass, the scale height, inclination, 
and PA of the major axis on the sky. 
The model is then convolved with the beam shape and spectral 
resolution profile and sampled at the observed pixel scales. 
The total mass and its radial distribution fully define the rotation curve, 
which is computed assuming the disk is entirely 
supported by ordered rotational motion (i.e. a thin disk).
An important parameter for the dynamical model is the elliptical 
beam shape. This is crucial for modelling marginally resolved sources, in order to 
properly account for beam smearing. Interferometric data, such those presented here, 
have a well characterized beam, which is listed in Table~1. 

DYSMAL can accommodate a departure from the ideal thin disk assumption, 
by allowing the thickness of the disk to be defined. 
If a finite scale height is specified, the associated thick disk kinematics can be derived in 2 ways.
With the first, the code makes an estimate of the local $\sigma$
based on the vertical 
height $H$ and rotation velocity $v$ at radius $R$, by setting $R/H = v/\sigma$ as for a compact disk.
Alternatively, once the geometry has been set, an additional fixed and uniform isotropic dispersion term can be included \citep{cre09,dav11}.

We have four observable constraints related to the kinematics: 
(\emph{i}) the integrated linewidth; 
(\emph{ii}) the major and (\emph{iii}) minor axis extent; 
and (\emph{iv}) the separation of the red and blue channel map peaks. 
We can use these to match a dynamical model to the data via a $\chi^2$ minimization.
The observed quantities in Table~\ref{tb:lines} can thus be compared with the output of our dynamical modelling, 
which is given in Tab~\ref{tb:model}. 

Due to the limited number of constraints, 
we need to make some reasonable simplifying assumptions as follows: 
(a) the gas exists in a rotating 
disk-like structure for which the inclination is the same as for the stars and/or 
for the H$_2$~(1-0)\,S(1), and the position angle of which is set by the orientation of the line joining the centroids of the red and blue channel maps 
(PA$_{r/b}$ in Tab~\ref{tb:lines}, mean values are used if two lines are available). 
These are given as the fixed parameters in Table~\ref{tb:model}. 
(b) The mass and light follow  the same distribution.
We use a Gaussian mass distribution, and derive the rotation curve from this. 
We note that assuming a uniform or exponential mass distribution makes no 
significant difference to the results.
(c) We can apply an axisymmetric model to emission that is not fully axisymmetric. 
In fact, significant velocity gradients due to unresolved non-circular motions can alter the kinematic symmetries. 
As shown by recent numerical simulations \citep{hop10}, the BH feeding process is related 
to gravitational instabilities that generically form lopsided, eccentric disks that propagate inwards from larger radii. 
This lopsided structures can dominate the disk kinematics within the central 10 pc \citep{hop12}.
We consider it unrealistic to try and allow the model to account for details in the observed flux distribution. 
Instead, to overcome this limitation, we ignore the fact that the centroids of the red and blue channel maps are 
not symmetrically placed either side of the continuum peak, and simply use their relative position. 

To match a model to the observed data, several of the input parameters in Table~\ref{tb:model} are left free to vary.
These are 
the \textit{intrinsic} width ($R$) and height ($H$) of the rotating disk together with 
the scaling of the rotation curve. The last ($M_{scale}$) is defined by specifying, for some radius, 
an enclosed mass, defined as the mass that is supported solely by ordered rotation (and which may therefore differ from the actual mass for a thick disk).
The accuracy of the centroid positions (and thus r/b separation) is, like the other uncertainties, 
estimated using Monte Carlo techniques. We create $10^3$ realizations of the model cube representing NGC~3227 in Table~5, 
each time randomly adjusting the data values according to the measured S/N in Table~\ref{tb:lines}, and re-fitting 
Gaussians. This yields $\pm0.02\arcsec$. We choose NGC~3227 for this test because it has the lowest S/N, 
and so for the other sources the uncertainties should be even smaller. 

To illustrate that the thin disk assumption is actually not appropriate to reproduce the observations, we show, 
in Table~5, 
several toy models for NGC~3227. 
The aim of these examples is not to fit the data (i.e. all the input parameters are not free to vary), 
but rather to demonstrate how the geometry and dynamics of the disk affect its observed properties.
After smoothing with the spatial beam, 
the size in each case is comparable to that of the data, but there are 
large discrepancies with the channel map separation and the linewidth. 
We start in model 1 with a FWHM of $0.5\arcsec$. 
From a comparison with the NGC~3227 data in Table~\ref{tb:lines}, the size is about right, but the channel map separation is rather too large and the linewidth much too small. 
We therefore address in model 2 whether the discrepancies might be resolved by increasing the rotation rate. 
To do this, we increase the rotationally supported mass of the first model by a factor of 2. 
While this does broaden the line as required, it still does not match the r/b channel separation, which remains too large. 
Instead, in model 3, we put the mass scaling back to its initial value, and instead consider a thick disk. 
To do so requires an additional intrinsic dispersion which broadens the line width; and in addition the beam smearing effects on the thick disk reduce the r/b separation.
The model yields a reasonable match for all the output parameters. 

Since a thick disk appears better suited to the data, we have left all 3 input parameters free to vary for each source. 
A comparison of the measurements in Table~\ref{tb:lines} with the model outputs in Table~\ref{tb:model}  
clearly shows that the models can match all the observed constraints; 
and that the large velocity dispersion of the molecular gas is associated with its geometric distribution. 
Indeed, for all of our sources, the dense gas exists at significant scale 
heights above the disk plane. 
Table~\ref{tb:kin} lists the \textit{intrinsic} quantities derived from models fitted to the dense gas kinematics.
It shows that the ratio H/R (and thus $\sigma/v$) is about 0.3, if we allow DYSMAL to estimate the dispersion associated with a given scale height. 
On the other hand, if we use a fixed uniform additional $\sigma$ term, we obtain $\sigma/v\sim0.5$. 
These two ratios are in approximate quantitative agreement, given the contrast between the simplicity of the models and greater complexity of the data. 
We thus adopt a mean $\sigma/v=0.4$ in the following discussion. 

The dynamical models allow us to compare the intrinsic dense gas kinematics with 
those observed for the warm molecular hydrogen and stars. 
The intrinsic velocity dispersions in Table~\ref{tb:kin} for NGC~3227 and NGC~4051 are about half of those measured from the warm gas, as given in Tab~3 of H09. 
The same conclusion is reached by comparison to the stellar dispersion in Table~4 of B06. 
As mentioned in Section~4 the difference can be reasonably ascribed to the 
high densities traced by \hcn\ and \hco. 
Nevertheless, the dense gas is still scattered to significant scale heights.

We note that it is also rotating faster than either the warm H$_2$ or stars.
Indeed the rotation 
velocity in our models at a radius of 30~pc is 105~km/s in NGC~3227 and 47~km/s NGC~4051, 
compared with 59~km/s and 37~km/s, respectively, measured from the H$_2$~(1-0)\,S(1) line (H09).

\subsection{Comparison with other galaxies}

In this section, we briefly compare the intrinsic dispersion of the dense gas in these four AGN as given in Table~\ref{tb:kin}, with that observed in two other extreme regimes.

The center of the Milky Way (MW) is a good choice for a reference of a quiescent galaxy because of its proximity.
Thus, even with modest angular resolution and sensitivity, one can detect \hcn\ on small physical scales and at low gas masses; and \hcn\ emission has been targeted numerous times. 
\cite{chr05} observed \hcn(1-0) 
and \hco(1-0) in the circumnuclear disk of the MW and resolved more than 20 molecular cores. 
The mean velocity dispersion they measure is $<\sigma>=11\pm0.4$~km/s, smaller than we find in our AGN. 
However, the MW circumnuclear disk is only 3~pc across, rather less than the size scales our data are probing. 
On more comparable scales of $\sim100$~pc in the MW, \cite{lee96} found typical values of $\sigma\sim10-21$~km/s, 
which is still smaller than in three of the four AGN. 
This suggests that there may be a heating source in the central regions of AGN that maintains the high dispersion in those objects, and which is not present in the MW.

The opposite regime is represented by strong star forming galaxies also harbouring AGN, 
such as Ultraluminous Infrared Galaxies (ULIRGs). In the three ULIRGs observed by \cite{ima06}, 
the \hcn(1-0) and \hco(1-0) lines show extremely large line widths, in the range 140-300~km/s FWHM. 
These are the directly observed integrated values, which should therefore be compared with our 80-200~km/s from Table~\ref{tb:lines}. 
The similar ranges suggest that the central regions of AGN may be as extreme as ULIRGs.
However, we note that the beam smearing correction is unknown for the ULIRG measurements, and since the data trace large spatial scales ($\sim1-3$~kpc), this may partially explain the high velocity dispersions observed. 

The implication of discrepant velocity dispersions in different regimes is intriguing, as 
it might reveal a link between the central disk structure and the nuclear star formation. 
This issue is discussed in the next section.

\section{Discussion}
\label{sec:discuss}

\begin{table}
 \begin{tabular}{lcccc}
 \hline

  Source         & R    & H   & $v$ & $\sigma$ \\
           &  pc  & pc  &  km/s & km/s   \\
 \hline
  NGC~2273 &  97  &  28 & 83   &   33    \\
 \hline
  NGC~3227 &  23  &  6  &  105 &  42    \\
 \hline
  NGC~4051 &  22  &  6  & 47   &  19    \\
 \hline
  NGC~6951 &   45 & 12 &  84  &  34      \\
 \hline
 \end{tabular}
 \caption{Intrinsic kinematics of the modelled disk. (1) Source name. 
 (2)-(3) Scaling radius and height given in parsecs. 
 (4)-(5) Rotation velocity and velocity dispersion at R. }

 \label{tb:kin}
\end{table}

We have found that in our four active galaxies the intrinsic \hcn\ and \hco\ velocity dispersion 
lies in the range $\sim20-40$~km/s. While this is about half of the velocity dispersion measured 
for the warm H$_2$ (H09), it is still remarkably high. 
Indeed, the high velocity dispersion of the gas implies that the dense gas distribution is geometrically thick.
But the mechanism responsible for maintaining it is unclear.
One inevitable conclusion is that such structures have to be clumpy, and the high dispersion of the clouds must be supported by a force other than thermal pressure, otherwise inelastic collisions should collapse the structure to a thin disk within an orbital timescale (a few Myr at our scale radii). 
Supernovae explosions in a post-starburst phase \citep{wad09} appear able to explain the lower end of the velocity 
dispersion range ($\sigma\sim20$~km/s).
On the other hand, the higher dispersions require an alternative process.
Gravitational turbulence induced by external gas accretion into the nuclear region \citep[hereafter V08]{vol08}, could represent a plausible way to explain values of 40~km/s, if this mechanism can be shown to work. 
Three phases characterize the V08 evolutionary model: first a short and massive gas inflow from the host generates 
a massive turbulent disk where star formation occurs. In a second phase supernovae explosion clear the intra-cloud 
medium leaving a geometrically thick disk dominated by dense gas cloud cores. 
Finally, when the mass accretion rate into the central regions decreases sufficiently, the disk becomes thin and transparent.
To assess the evolutionary phase characterizing our sources, 
we first estimate whether or not the modelled disks are stable against star formation, 
and we then investigate the SF properties in their nuclear regions.

\subsection{Can stars form in the dense gas?}

\begin{table}
 \begin{tabular}{lcccc}
 \hline
 Source   & M$_{dyn}$    &   L$_{\hcn}$                   & $f_{gas}$ & Q/Q$_c$ \\
          & $10^7$ M$\odot$ & $10^6$ K km s$^{-1}$ pc$^2$   & \%        &         \\   
 \hline
 NGC 2273 & 23         & $3.2\pm0.1$                   & 14        & 4.2 \\
 \hline
 NGC 3227 & 9.7         & $0.59\pm0.08$              & 7         & 8.0 \\
 \hline
 NGC 4051 & 1.7       & $0.12\pm0.04$               & 7         & 8.1 \\
 \hline
 NGC 6951 & 11         &  $2.45\pm0.05$             &  22        & 2.5  \\
 \hline
 \end{tabular}
 \caption{Gas properties within the scaling radius R. 
 (1) Dynamical mass M$_{dyn}=(v^2+3\sigma^2)$R/G (see text for details). 
 Excluding the uncertainty of the coefficient  for the $\sigma$ term, random errors on dynamical masses are $15-20\%$. 
 (2) \hcn\ luminosity, which has been found to be proportional to the dense gas mass (Eq.\ref{lhcn}, \citealt{gao04}, \citealt{kri08}). 
 (3) Gas fraction, for which the relative error is in the range $20-30\%$. 
 (4) Toomre Q parameter normalized to the critical value of 0.68 \citep{dek09}.}
 \label{tb:qpar}
\end{table}

A rotating gaseous disk becomes unstable once the local gravity overcomes both differential 
rotation and turbulent pressure, i.e. when the Toomre Q parameter \citep{too64} is smaller than 
a critical value Q$_c$: 
\begin{center}
\begin{equation}
Q = \frac{\kappa\sigma}{\pi G \Sigma_{gas}} < Q_c.
\label{Q} 
\end{equation}
\end{center}
In Equation~\ref{Q}, G is 
the gravitational constant, $\sigma$ the local velocity dispersion; 
and $\kappa$ is the epicyclic frequency, which is a function of the angular velocity $\Omega$. 
An isothermal thick disk has a critical Q parameter $Q_c=0.68$ \citep{dek09}. 
For a uniform disk the epicyclic frequency becomes:
\begin{equation}
\kappa = \sqrt{3}\Omega = \sqrt{3} v/R,
\label{k} 
\end{equation}
with $v$ the rotation velocity at radius $R$. 
The surface density of the gaseous disk in Eq.~\ref{Q} is: 
\begin{equation}
\Sigma_{gas} = \frac{M_{gas}}{\pi R^2}=\frac{f_{gas} M_{dyn}}{\pi R^2},
\label{s} 
\end{equation}
where, in the right part, we replace the gas mass M$_{gas}$ with the derived dynamical 
mass M$_{dyn}$ scaled by the gas fraction $f_{gas}$. 
Because of significant random motions ($\sigma/v \sim0.4$ from Table~\ref{tb:kin}), we note that M$_{dyn}$, in turn, 
depends on both rotation velocity and dispersion:
\begin{equation}
M_{dyn} = (v^2+3\sigma^2)R/G.
\label{md} 
\end{equation}
Equation~\ref{md} arises from combining the kinetic energy of the random motions, 
which we take to be $3\sigma^2$ (assuming $\sigma$ comes from macroscopic turbulence, and that we measure a one-dimensional component of an isotropic three-dimensional distribution), 
with a $v^2$ term for the rotation. 
Finally, we use the relation between size scales and kinematics for a compact disk, which gives: 
\begin{equation}
\sigma= v H/R.
\label{th}
\end{equation}
By substituting Equations~\ref{k}, \ref{s}, \ref{md}, and \ref{th} into the Q parameter definition (Eq.~\ref{Q}) 
we obtain a simple expression for Q, that depends only on the geometric structure 
of the thick disk and the gas fraction:
\begin{equation}
Q = \sqrt{3} \frac{H}{R} \frac{1}{(1+3H^2/R^2)f_{gas}}.
\label{qdf}
\end{equation}
We note that Eq.~\ref{qdf} is, for example, analogous to the expression derived in \cite{gen11}, 
but with an added term corresponding to the mass supported by random motions.

It is straightforward to obtain the dense gas fraction $f_{gas}=M_{gas}/M_{dyn}$, once the molecular 
gas mass is estimated. 
The \hcn\ luminosity (L$_{\hcn}$) is related to $M_{gas}$ through \citep{gao04,kri08}:
\begin{equation}
M_{gas} = 10 L_{\hcn}~M_\odot~(K km s^{-1} pc^2)^{-1}. 
\label{lhcn}
\end{equation}
The dynamical mass, as given in Eq.~\ref{md}, is listed in Table~\ref{tb:qpar} together with the \hcn\ luminosity, 
gas fraction and the Toomre Q parameter computed at the scale radius R taken from Table~\ref{tb:kin}.
The estimate of the uncertainties given in the Table is based only on the propagation of random errors, 
and does not include systematic uncertainties such as the coefficients used to estimate the dynamical mass 
or the uniqueness of the relation between HCN luminosity and gas mass. 
Nevertheless, with a gas fraction of $\sim10-20\%$ as derived above, there are indications that 
the Q parameter is greater than the critical value, which would imply that the central 30-100~pc of our AGN are stable against star formation.
 
\begin{figure*}
\centerline{
\hbox{
\includegraphics[width=0.79\linewidth,angle=0]{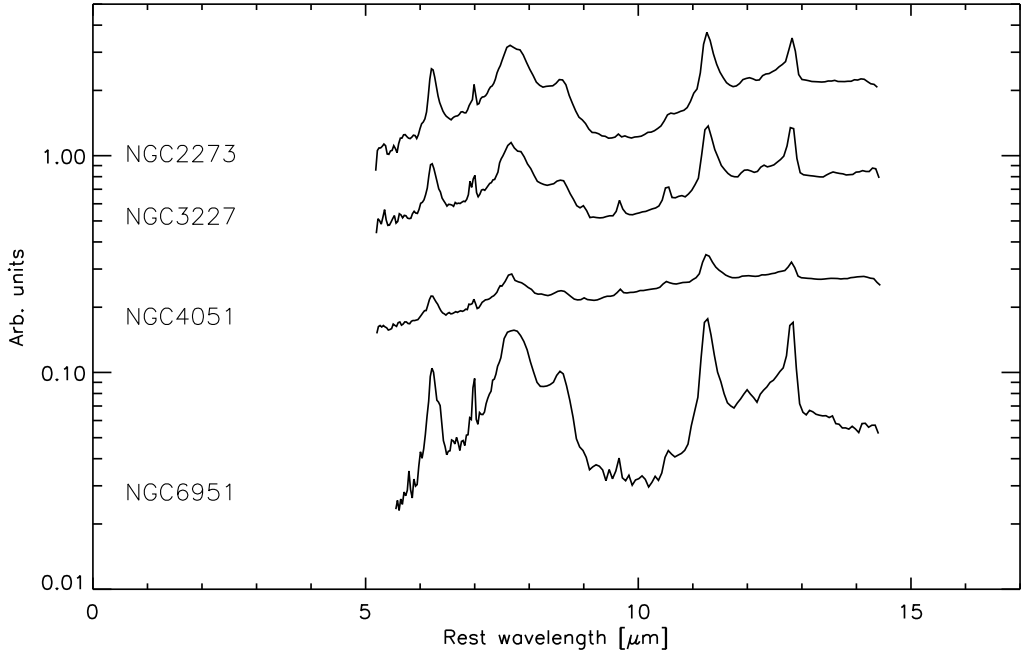}
}
}
\caption{\emph{Spitzer}/IRS spectra for the four AGN. 
The spectra are arbitrarily scaled for clarity, and are ordered with decreasing AGN luminosity from the top. 
The data reduction is described in \citet{san10}. 
The four spectra show typical PAH emission features 
at 6.2~$\mu$m, 7.7~$\mu$m, 8.6~$\mu$m, 11.3~$\mu$m, 12.7~$\mu$m and 
have been previously analysed by \citet{dia10} (NGC~2273), 
\citet{mel08} (NGC~3227), \citet{san10} (NGC~4051), and \citet{dia10} (NGC~6951).}
\label{fg:spitzer}
\end{figure*}

\subsection{Is there evidence for star formation?}

The high Toomre Q parameters derived for our sample (Table~\ref{tb:qpar}), indicate 
that it should not be possible for the gas to rapidly form stars. 
We expect, therefore, that no significant star formation should be detected in the central $\sim100$\,pc regions observed with the PdBI. 
To verify this conjecture, we look for ongoing SF by investigating mid-infrared spectra in the context of other observations. 
The $5-15~\mu$m waveband is expected to show strong polycyclic aromatic hydrocarbon 
(PAH) features if the molecules are excited by 
radiation from young massive stars \citep{all89}.
The mid-IR spectra obtained with \emph{Spitzer}/IRS \citep{hou04} are shown in Fig.~\ref{fg:spitzer}. 
All the sources exhibit strong PAH emissions, suggestive of significant star formation.
But it is not clear where the SF is spatially located. 
Indeed, due to the IRS slit aperture of $3.6\arcsec$, the spectra are integrated over the central 
450~pc, 305~pc, 180~pc and 370~pc respectively for NGC~2273, NGC~3227, NGC~4051 and NGC~6951.
Data obtained at high angular resolution can help, but it can still sometimes be hard to reach a firm conclusion. 

As described in Section~3.2, in NGC~3227 the \hcn\ nuclear emission is surrounded by an annular structure 
seen with CO(2-1) line and H-band continuum observations (Fig.~\ref{fg:maps}, \citealt{sch00a}, \citealt{dav06}). 
This circumnuclear ring hosts the bulk of the current SF; while at radii $<30$~pc, SF has recently ceased. 
This has been shown by adaptive optics observations \citep{dav06}, which reveal that the Br$\gamma$ equivalent width is inconsistent with on-going SF.
Based on this and other diagnostics, these authors concluded that the inner tens of parsecs in NGC~3227 have experienced a short starburst event in the last 40~Myr, which has now ceased.

For both this source and NGC~2273, B06 find a reduction in the stellar velocity dispersion 
across the ring, which they ascribe to recent SF in dynamically cool gas.
The implication is that the higher dispersion in the nucleus indicates that SF is not active there. 
For NGC~2273, the circumnuclear ring is clearly visible also in HST/NICMOS 
images \citep{mar03}. 
From their spectroscopic data, \cite{fer00} relate H$\alpha$ emission to HII regions in the ring, whereas they argue that the nuclear [OIII]/H$\alpha$ ratio is typical of AGN. 
Thus, although we cannot firmly rule it out, these results from B06 and \cite{fer00} suggest that, 
as for NGC~3227, there is no current SF in the nucleus. 

In NGC~4051 stellar kinematic measurements have been prevented by the strong AGN emission 
(B06; \citealt{rif08}, hereafter R08), 
and the central physics are more puzzling. The Br$\gamma$ emission is very compact and it is hard to 
draw firm conclusion: R08 find no rotation and argue that most of the emitting gas is not 
confined to the plane of the galaxy; in contrast, \cite{mue11} trace a Br$\gamma$ 
velocity field consistent with rotation, that is seen also in the H$_2$ 1-0\,S(1) and stellar kinematics (H09). 
From the H$_2$ line ratios R08 argue that the gas is thermalised rather than fluorescently 
excited as one might expect from recent star formation.
Although this could instead be due to high densities, the H$_2$~(1-0)\,S(1)/Br$\gamma$ ratio in the nucleus is about 1, higher than expected if the heating process is due to UV radiation from young stars. 
Moreover, R08 also detect [OIII] and [CaVIII] lines characterized by blue wings 
and argue that significant line emission originates from the AGN narrow-line region. Thus, a similar 
origin of the compact Br$\gamma$ emission is plausible.
On the other hand, if we do attribute the Br$\gamma$ emission to SF, we can estimate its equivalent width once the flux measured by \cite{mue11} is corrected for dilution (see Fig.~1 in R08). 
This yields an equivalent width of about 7\AA, a low value for SF that, according to 
\cite{dav07}, implies ceased SF activity. 
Finally, \cite{rod03} are able to give only an upper limit for the $3.3~\mu$m 
PAH feature using a $0.8\arcsec$ slit. We can thus conclude that the SF is NGC~4051 is mainly located 
between $0.8\arcsec$ and $3.6\arcsec$, and if a nuclear component within the central 40~pc is present, 
it is not significant. 

In addition to its AGN, NGC~6951 also exhibits a pronounced starburst 
ring located at $5\arcsec$ (480~pc) from the nucleus. 
Plausibly, most of the PAH emission may originate from this ring, which has been observed in 
several wavebands: near-IR \citep{mar03}, 
optical (e.g. \citealt{sto07} and references therein),  and radio \citep{sai02}.
High resolution observations detect strong CO and \hcn\ emission both in the ring and 
the central AGN (K07; \citealt{gar05}, \citealt{koh99}).  
While the \hcn/CO ratio in the ring is typical of a starburst, 
the \hcn\ is significantly enhanced in the nucleus. The higher nuclear ratio could be due  
either to denser and/or hotter gas than in the starburst ring, 
or because the gas chemistry in the nucleus of NGC 6951 is dominated by X-ray radiation
 from the AGN, skewing relative abundances (see \citealt{dav12}).
Thus, again, significant SF close around the AGN cannot easily be excluded. 
The radio observations can help to address this issue. The supernova rate, 
estimated from the nuclear radio flux density, is $\sim0.003$~yr$^{-1}$ \citep{sai02}, much 
smaller than would be expected for strong ongoing SF, and rather less than in NGC~3227 where the SF has ceased \citep{dav06}.
Indeed, if the SF is continuous, we can estimate an associated SF rate (SFR) from Equation~20 in \cite{con92},
obtaining SFR(M$>1$M$_\odot$)$\sim0.15$~M$_\odot$yr$^{-1}$ 
(including stars below 5~M$_\odot$, thus a factor of 2 higher as for the original \cite{con92} relation). 
This is comparable to the time-averaged SFR for the nuclear regions of other AGN \citep{dav07},
and provides an upper limit to the actual on-going SFR.

\section{Conclusions}

We have analysed 3~mm interferometric data obtained with the PdBI tracing the \hcn(1-0) 
and \hco(1-0) molecular lines in four local AGN. Our main conclusion, that these lines typically trace a thick disk with suppressed star formation, is based on the following: 

1. The line emission is marginally resolved in all four Seyfert galaxies.
When data for both \hcn\ and \hco\ are available, their spatial extents are in good agreement within the errors. 

2. The global kinematics are consistent with rotation.
We create dynamical models that match the observations, using four constraints, namely the integrated linewidth, the diameters of the major and minor axes, and the separation of 
the red and blue channel map centroids. The resulting structure is a geometrically thick disk.

3. The most remarkable feature is a high velocity dispersion that, once corrected for beam 
smearing using the dynamical models above, lies in the range 20-40~km/s. This is about a factor 2 lower than the stellar and/or H$_2$ 1-0\,S(1) 
dispersion, but is still remarkably high considering that we are tracing 
dense molecular gas. It implies a vertical height H$\sim30\%$ of the scale radius R.

4. Finally, we estimate the gas fraction within R ($f_{gas}\sim10-30\%$) and the Toomre Q parameter, 
finding that the central $\sim100$~pc should be stable against star formation.
This result appears to be observationally confirmed by the lack of ongoing nuclear SF in the nucleus (although there is plentiful evidence for on-going star formation in circumnuclear rings).
NGC~4051 remains the most puzzling source, with the lowest velocity dispersion and unclear evidence for or against nuclear star formation. 


\section*{Acknowledgements}
We thank the anonymous referee for a careful and thorough review of the manuscript, and for providing comments that have helped to improve it. 
E. Sani acknowledges financial support from ASI under Grant I/009/10/0/.

\label{lastpage}

\end{document}